%% file: main.tex
\pdfoutput=1
\documentclass[10pt,twocolumn]{article}
\usepackage[margin=1in]{geometry}
\usepackage{amsmath,amssymb,amsthm,bm,mathtools}
\usepackage{graphicx}
\usepackage{booktabs}
\usepackage{microtype}
\usepackage[numbers,sort&compress]{natbib}
\usepackage[colorlinks,linkcolor=blue!60!black,citecolor=blue!60!black,urlcolor=blue!60!black]{hyperref}
\usepackage{xcolor}
\newtheorem{theorem}{Theorem}
\newtheorem{proposition}{Proposition}
\newcommand{\ka}{\mathbf{k}}
\newcommand{\Tr}{\mathrm{Tr}}
\newcommand{\Dd}{\Delta_{\mathrm d}}
\newcommand{\Dz}{\Delta_0}
\newcommand{\Gt}{\hat G_2}
\newcommand{\cA}{\mathcal A}
\newcommand{\cC}{\mathcal C}
\newcommand{\cD}{\hat{\mathcal D}}
\newcommand{\cK}{\mathcal K}

\title{\bfseries Chiral superconductivity on a closed Klein bottle:\\ a chirality wall and its Majorana seam spectrum}
\author{Junhao (Eigen) Wang\\[2pt] \small Department of Mathematics, New York University, New York, NY, USA\\ \small \texttt{jw10111@nyu.edu}}
\date{\small Preprint --- 5 November 2026}

\begin{document}
\maketitle

\begin{abstract}
We study a spinless $\pi$-flux lattice superconductor whose normal-state momentum-space fundamental domain is a Klein bottle. Within this model, a glide lift and a hidden antiunitary symmetry independently force the Bogoliubov--de Gennes (BdG) Chern number to vanish. A diagonal pairing breaks the antiunitary obstruction exactly when its real part is nonzero; this is necessary, not sufficient, for a chiral phase. We derive explicit Pfaffians determining Chern parity and identify three analytic families of gap closings in a real-pairing section. For a chiral state on a closed real-space Klein bottle, the prescribed pairing and gluing produce a $\Delta C=2$ chirality wall. A translation-invariant unfolding resolves two co-propagating Majorana branches. Their finite-size levels depend on circumference and holonomy; the near-zero pair is lifted at first order by allowed local seam perturbations. A cut-cylinder control has the same low-energy channel count, so gluing reconnects existing gapless content. We also distinguish literal gluing from deck projection: particle--hole symmetry exchanges the deck eigenblocks, precluding their interpretation as independent BdG quotient Hamiltonians. Finite odd-frequency correlations track holonomy and local splitting, while cylinder and ordinary-wall controls show that their presence is not specific to Klein gluing. The contribution is a model-specific account of pairing compatibility, closed-surface seam spectra and finite correlations. The energy cost is evaluated for a fixed pairing background; neither a protected zero mode nor a self-consistent wall solution is claimed.
\end{abstract}

\section{Introduction}\label{sec:intro}

A $\mathbb Z_2$ gauge field on a lattice can make the Brillouin zone non-orientable. Chen, Yang and Zhao showed that with $\pi$ flux per plaquette the crystalline symmetries obey a projective algebra under which the momentum-space fundamental domain becomes a Klein bottle, and that the topological classification on such a Brillouin Klein bottle is $\mathbb Z_2$ rather than the Chern number $\mathbb Z$ \cite{ChenYangZhao2022}. The idea has since been realized in classical-wave lattices and extended in several directions, most recently to twisted Brillouin Klein bottles, whose covering torus can carry a nonzero (even) Chern number \cite{XiaoZhao2026}. In parallel, real-space Klein bottles have entered lattice field theory: Seiberg and Zhang placed $2{+}1$d staggered fermions---the same $\pi$-flux model---on sheared tori and Klein bottles to diagnose mod-8 parity/time-reversal anomalies \cite{SeibergZhang2026}, and Araki \emph{et al.}\ formulated the $\mathbb Z_8$-valued Arf--Brown--Kervaire invariant of lattice Majorana fermions on non-orientable lattices \cite{ArakiABK2025}. Non-orientable manifolds have long been the natural probes of time-reversal and reflection anomalies \cite{Witten2016parity,Witten2016RMP} and of topological order twisted by parity \cite{ChanTeoRyu2016}.

Chiral superconductivity on a non-orientable surface has been studied before, in the open geometry of a M\"obius band: Quelle, Kvorning, Hansson and Morais Smith showed that a chiral $p$-wave state on a M\"obius band must contain a chirality domain wall, that the wall and the edge each carry a Majorana zero mode without any flux, and that twisted boundary conditions select the symmetric zero mode \cite{Quelle2016}; the same group had earlier analyzed the Chern obstruction and the double-cover method on the M\"obius band for quantum Hall states \cite{Beugeling2014}, and the domain wall between opposite $p$-wave chiralities is known to carry a pair of co-propagating Majorana channels \cite{Serban2010}. Interacting Klein-bottle geometries have been used to define many-body invariants of topological insulators and superconductors \cite{SonAlicea2019}. The question we ask here is what a gapped chiral superconductor does on a \emph{closed} Klein bottle, i.e.\ without any edge, when the model is a definite $\pi$-flux lattice Hamiltonian whose momentum space is itself a Klein bottle. The orientation obstruction---a Chern number is a pseudoscalar---requires a wall if the uniform gapped chiral bulk is retained on both sides and reconstruction is confined to a finite seam width \cite{Beugeling2014,Quelle2016}. This is the fixed pairing background considered here; a bulk gap closing or loss of the order parameter is a different resolution. The obstruction does not decide the finite-lattice questions: which pairings are compatible with the gluing, what spectrum the wall carries when it is the only defect on the surface, whether the resulting near-zero modes are protected, and what the non-orientable gluing adds to the domain-wall physics already present on a cylinder.

Odd-frequency pairing provides a complementary way to examine the same finite spectrum. Interface-induced pair correlations and their relation to Andreev and Majorana states are established subjects \cite{TanakaSatoNagaosa2012,TanakaTamuraCayao2024}; odd-frequency enhancement can also accompany trivial Andreev states \cite{AhmedTamuraTanakaCayao2025}, and multicomponent lattices can have a bulk odd-frequency background \cite{TamuraSSH2020}. We therefore ask how the anomalous Green function records this model's gluing, holonomy and local splitting, using the cut cylinder, uniform torus and an ordinary two-wall torus as controls. These are finite imaginary-frequency diagnostics of a prescribed static pairing background, not a new order parameter or a sufficient test for a topological zero mode.

Our results are as follows. (i) In momentum space, the glide $G_2$ that defines the Brillouin Klein bottle confines the BdG Chern number to $C=0$ (Theorem~1, an elementary statement we record in the form $C=\chi(S)\det(\mathrm Dg)\,C$). The confinement is enforced by two distinct operators in the pairing channel: the unitary glide lift $\Gt=i\tau_z\otimes\sigma_x$, which for $p+ip$ pairing is a symmetry only when accompanied by a $\pi/2$ rotation of the charge $U(1)$, and a hidden antiunitary, orientation-preserving $\cA=(\tau_z\otimes\sigma_x)\cK$. A diagonal pairing $\Dd$ breaks $\Gt$ for any phase but breaks $\cA$ if and only if $\mathrm{Re}\,\Dd\neq0$; this is a \emph{necessary} condition for $C\neq0$. The parity of $C$ is fixed by the Pfaffians of $H(\mathbf K)\tau_x$ at the four particle--hole-invariant momenta (Prop.~\ref{prop:pf}), which we verify pointwise on the phase diagram (Fig.~\ref{fig:phase}); the gap closings of the $(\mu,\mathrm{real}\,\Dd)$ section that we identify form three analytic families---the Pfaffian ellipses, the lines $\mu=\pm2\sqrt2\,t$, and a nodal segment on $\Dd=\pm\Dz$, $2t\le|\mu|\le2\sqrt2\,t$---and there are gapped $C=\pm2$ phases. (ii) The projective symmetry algebra of the glide-symmetric BdG model closes into sixteen elements, eight unitary and eight antiunitary; its unitary part satisfies the quaternion relations $i^2=j^2=k^2=ijk=-1=(-1)^F$ with $i=\Gt$, $j=\cC\cA$, $k=-i\tau_x$, but only $\pm1$ and $\pm\Gt$ are symmetries of a fixed Hamiltonian---$j$, $k$ and their multiples are spectral antisymmetries ($H\to-H$)---and the charge $U(1)$ is a covariance of the pairing family, not a symmetry. (iii) On a real-space Klein bottle $(x,y+L_y)\sim(1-x,y)$, the glide-covariant $p+ip$ state commutes with a deck lift $\cD$ satisfying $\cD^2=(-1)^F$, whose two eigenblocks are isospectral and gapped; we show that these blocks are exchanged by particle--hole conjugation and are therefore \emph{not} independent BdG Hamiltonians on the quotient, and that none of the sixteen literal gluings with a $c$-number bond phase reproduces them. The $C=-1$ state has no uniform deck lift: the gluing produces a $\Delta C=2$ chirality wall along the seam. (iv) The wall carries two co-propagating chiral Majorana branches, degenerate throughout the resolved in-gap window, which we resolve on a translation-invariant unfolding of the seam and compare with a torus carrying two hand-made walls and with the cylinder obtained by cutting the seam bonds; the seam levels $2t\cos q$ at the allowed momenta with their dependence on circumference and $x$-holonomy, the near-zero pair and its exponential $L_y$ dependence, and a ledger of local seam perturbations classified by amplitude scaling are given. (v) Finite odd-frequency correlations are compared in the same holonomy and perturbation sectors. Their low-energy spectral contribution and ordinary-wall control delimit what can be attributed to the Klein seam. (vi) The seam tension, the entanglement log-coefficient and a real-space symmetry search over a large family of candidate operations quantify what the gluing costs and what it does not protect.

The contribution is model specific. The collapse $\mathbb Z\to\mathbb Z_2$ on the Brillouin Klein bottle is due to Ref.~\cite{ChenYangZhao2022}; that an untwisted Klein bottle forces $C=0$ is stated in Ref.~\cite{XiaoZhao2026}; the magnetic translation algebra and the Klein-bottle quotient of staggered fermions are in Ref.~\cite{SeibergZhang2026}; the forced chirality wall and its flux-free Majorana modes on a M\"obius band are in Refs.~\cite{Beugeling2014,Quelle2016}; the relation between the Chern parity of a class-D superconductor and Pfaffians at particle--hole-invariant momenta is standard \cite{Sato2010}. What is new is the superconducting specialization of the Brillouin Klein bottle (the two obstructions, the $\mathrm{Re}\,\Dd$ criterion and the phase diagram of the $(\mu,\mathrm{real}\,\Dd)$ section with its three analytic node families), the spectral symmetry algebra, and the closed-surface seam: its momentum-resolved anatomy, its holonomy and perturbation dependence, and its tension. The near-zero pair is not protected, and the deck eigenblocks are not independent BdG quotient Hamiltonians. These distinctions delimit the physical interpretation of the seam. The added correlation analysis links its finite spectrum to a familiar pairing diagnostic; neither odd-frequency pairing itself nor its appearance near an interface is claimed as a new mechanism.

The paper is organized as follows. Section~\ref{sec:model} sets up the model. Section~\ref{sec:confine} proves Chern confinement, identifies the two obstructions, and gives the phase diagram. Section~\ref{sec:algebra} gives the symmetry algebra. Section~\ref{sec:KB} places the model on a real-space Klein bottle and dissects the seam. Section~\ref{sec:cost} quantifies the entanglement and the energetic cost of the seam. Section~\ref{sec:discussion} discusses what is established and what is not. Appendices collect methods and cross-checks, the normal-state Klein-bottle $\mathbb Z_2$ invariant, the glide-bilayer control, finite odd-frequency definitions and derivations, and the reproduction package. A separate supplement gives the full four-channel correlation plots and finite comparison tables.

\section{Model and the Brillouin Klein bottle}\label{sec:model}

\begin{figure*}[!t]
\centering\includegraphics[width=0.96\textwidth]{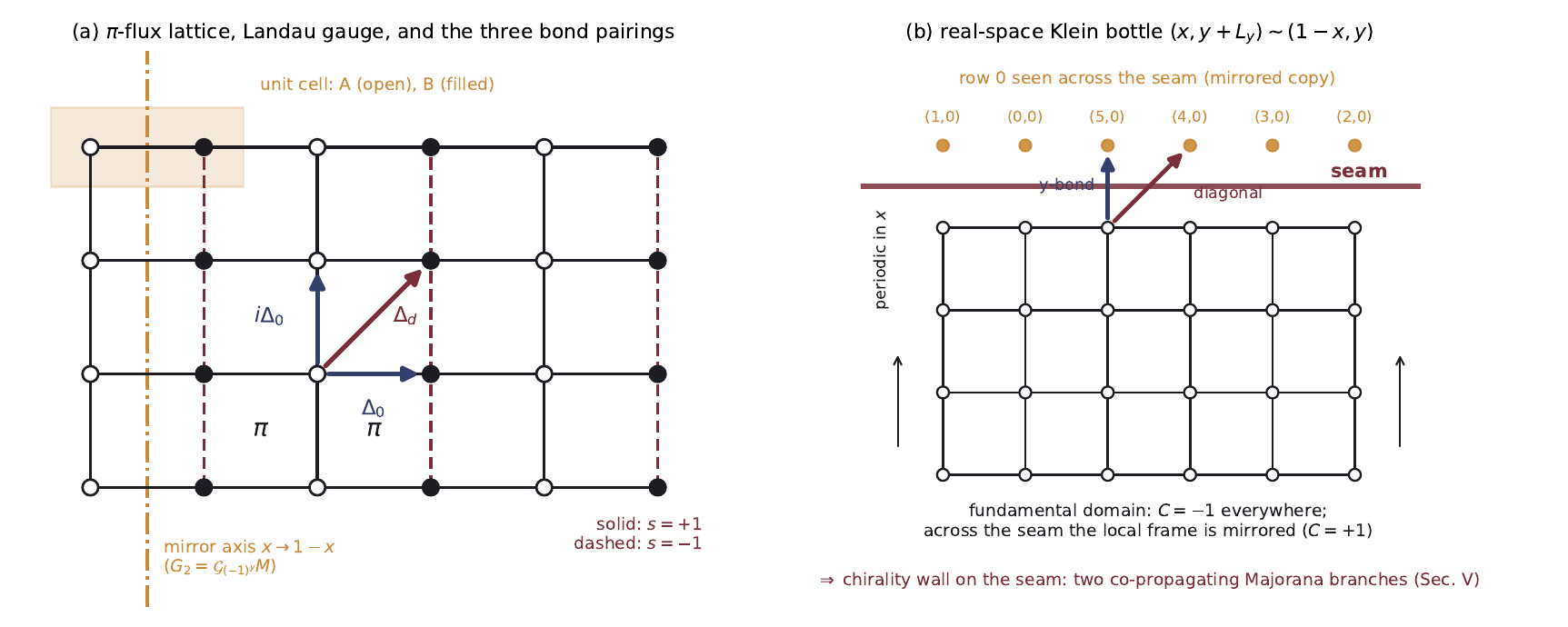}
\caption{The model. (a) Square lattice with $\pi$ flux per plaquette in Landau gauge (horizontal bonds $s=+1$, vertical bonds $s=(-1)^x$, dashed $=-1$); the two-site unit cell $A/B$; the three odd-parity bond pairings of Eq.~\eqref{eq:H} drawn from one site ($\Dz$ on $\hat x$, $i\Dz$ on $\hat y$, $\Dd$ on $\hat x+\hat y$); and the bond-centered mirror axis $x\to1-x$ whose combination with the gauge transformation $(-1)^y$ is the real-space form of the glide $G_2$. (b) The real-space Klein bottle of Sec.~\ref{sec:KB}: the fundamental domain $L_x\times L_y$ is periodic in $x$, and its top row is glued to a mirrored copy of its bottom row, $(x,L_y)\sim(1-x,0)$; a cross-seam $y$ bond and a cross-seam diagonal bond are shown. For chiral pairing the local frame across the seam is mirrored ($C=-1\to C=+1$), so the seam is a chirality wall.}
\label{fig:lattice}
\end{figure*}

\begin{figure*}[!t]
\centering\includegraphics[width=0.66\textwidth]{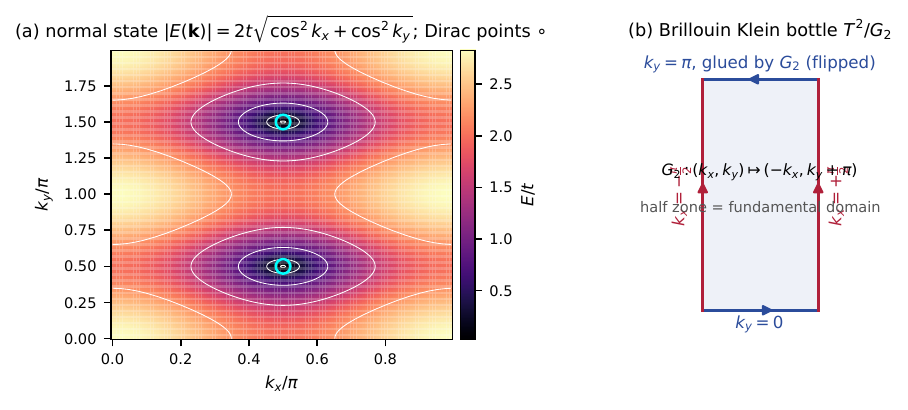}
\caption{Normal state. (a) The two-band $\pi$-flux spectrum $|E(\ka)|=2t\sqrt{\cos^2k_x+\cos^2k_y}$ on the half Brillouin zone $k_x\in[0,\pi)$ with the two Dirac points at $(\pi/2,\pi/2)$ and $(\pi/2,3\pi/2)$. (b) The Brillouin torus modulo the orientation-reversing free glide $G_2:(k_x,k_y)\mapsto(-k_x,k_y+\pi)$ is a Klein bottle.}
\label{fig:model}
\end{figure*}

We consider spinless fermions on the square lattice with a static $\mathbb Z_2$ gauge field $s_{ij}=\pm1$ on the bonds,
\begin{align}
H&=\sum_{\langle ij\rangle}\big[-t\,s_{ij}\,c_i^\dagger c_j+\mathrm{h.c.}\big]-\mu\sum_i n_i\nonumber\\
&\quad+\sum_i\Big[\Dz\,c_i^\dagger c_{i+\hat x}^\dagger+i\Dz\,c_i^\dagger c_{i+\hat y}^\dagger\nonumber\\
&\qquad\qquad+\Dd\,c_i^\dagger c_{i+\hat x+\hat y}^\dagger+\mathrm{h.c.}\Big],
\label{eq:H}
\end{align}
with $\pi$ flux per plaquette, $\prod_{\partial\square}s=-1$ (Fig.~\ref{fig:lattice}a). In Landau gauge (horizontal bonds $+1$, vertical bonds $(-1)^x$) the unit cell doubles along $x$; with cell vectors $(2,0)$ and $(0,1)$ and the Bloch phase $e^{i(2k_xR_x+k_yR_y)}$ per cell separation, the normal-state Bloch Hamiltonian is
\begin{equation}
\begin{split}
H_0(\ka)={}&-2t\cos k_y\,\sigma_z-t(1+\cos2k_x)\,\sigma_x\\
&-t\sin2k_x\,\sigma_y,
\end{split}
\label{eq:H0}
\end{equation}
with spectrum $E=\pm2t\sqrt{\cos^2k_x+\cos^2k_y}$, $k_x\in[0,\pi)$, $k_y\in[0,2\pi)$, and two Dirac cones at $(\tfrac\pi2,\tfrac\pi2)$ and $(\tfrac\pi2,\tfrac{3\pi}2)$ (Fig.~\ref{fig:model}). (The overall sign of Eq.~\eqref{eq:H0} is that of the code; it is immaterial for everything below, but a sign flip of $H_0$ alone is not a symmetry once pairing is added, so we keep the code convention throughout.) This is the $2{+}1$d staggered fermion of Ref.~\cite{SeibergZhang2026}. It has the momentum-space glide
\begin{equation}
\begin{gathered}
G_2:(k_x,k_y)\mapsto(-k_x,k_y+\pi),\\
\sigma_xH_0(\ka)\sigma_x=H_0(G_2\ka),
\end{gathered}
\end{equation}
an orientation-reversing free involution of the Brillouin torus, so that the fundamental domain $T^2/G_2$ is a Klein bottle \cite{ChenYangZhao2022}. In real space $G_2$ is the bond-centered mirror $x\to1-x$ combined with the gauge transformation $(-1)^y$.

Pairing enters through the Nambu spinor $\Psi_\ka=(c_{\ka A},c_{\ka B},c^\dagger_{-\ka A},c^\dagger_{-\ka B})^{\mathsf T}$ and the BdG Hamiltonian
\begin{equation}
H_{\rm BdG}(\ka)=\begin{pmatrix}H_0(\ka)-\mu&\Delta(\ka)\\ \Delta^\dagger(\ka)&-H_0^{\mathsf T}(-\ka)+\mu\end{pmatrix},
\end{equation}
$\Delta(\ka)=D(\ka)-D^{\mathsf T}(-\ka)$ for the directed bond amplitudes $D$, with particle--hole conjugation $\cC=\tau_x\cK$, $\cC H_{\rm BdG}(\ka)\cC^{-1}=-H_{\rm BdG}(-\ka)$ (class D). Since the fermions are spinless, an equal-time on-site pairing term vanishes and all three pairings in Eq.~\eqref{eq:H} are odd-parity bond pairings. This does not exclude unequal-time on-site anomalous correlations; their frequency parity follows from fermionic exchange (Appendix~\ref{app:odd}). Throughout, ``gap'' means $E_g=\min_\ka|E_{\rm BdG}(\ka)|$ and $t=\Dz=1$ unless stated. A vortex in a class-D superconductor with Chern number $C$ binds Majorana modes of which $C\bmod2$ are protected \cite{ReadGreen2000,FuKane2008,TeoKane2010}; the criterion of interest is therefore \emph{odd $C$}.

\section{Chern confinement, the two obstructions, and the phase diagram}\label{sec:confine}

The two Dirac cones sit at $k_y=\pi/2$ and $3\pi/2$, and a $p+ip$ gap function with an $i\sin k_y$ component takes opposite signs on them, so the two cones acquire opposite masses and the total Chern number vanishes wherever the state is gapped. This is not an accident of $p+ip$ but a consequence of the glide, as the following elementary statement makes precise.

\begin{theorem}[Chern confinement]\label{thm:confine}
Let $H(\ka)$ be a gapped Bloch (BdG) Hamiltonian on the Brillouin torus and $S$ an operator with $S\,H(\ka)\,S^{-1}=H(g\ka)$ for an affine diffeomorphism $g$ of $T^2$, mapping the occupied (negative-energy) space at $\ka$ to the occupied space at $g\ka$. Let $\chi(S)=+1$ ($-1$) for unitary (antiunitary) $S$. Then the total Chern number of the occupied bands obeys
\begin{equation}
C=\chi(S)\det(\mathrm Dg)\,C.
\end{equation}
In particular $C=0$ whenever $\chi(S)\det(\mathrm Dg)=-1$.
\end{theorem}
\begin{proof}
With $P(\ka)$ the occupied projector and $\omega=\tfrac{i}{2\pi}\Tr(P\,\mathrm dP\wedge\mathrm dP)$ the gauge-invariant Berry curvature, $P(g\ka)=SP(\ka)S^{-1}$ gives $g^*\omega=\omega$ for unitary $S$ and, because $\Tr(P\,\mathrm dP\,\mathrm dP)$ is purely imaginary, $g^*\omega=-\omega$ for antiunitary $S$. The change-of-variables formula gives $\int g^*\omega=\mathrm{sgn}\det(\mathrm Dg)\int\omega$.
\end{proof}

\noindent The four quadrants are: unitary orientation-preserving---no constraint; unitary orientation-reversing (mirror, glide)---$C=0$; antiunitary orientation-preserving (time-reversal type)---$C=0$; antiunitary orientation-reversing---no constraint. Particle--hole conjugation is antiunitary and orientation-preserving in two dimensions but anticommutes with $H$, mapping occupied to empty states; it is outside the hypothesis and yields only $C_{\rm empty}=-C_{\rm occ}$, which is why class D retains a $\mathbb Z$ classification. The theorem is the band-theoretic form of the continuum fact that a Chern--Simons level needs an orientation \cite{Witten2016parity}; that an untwisted Brillouin Klein bottle forces $C=0$ is stated in Ref.~\cite{XiaoZhao2026} following Ref.~\cite{ChenYangZhao2022}, and Ref.~\cite{XiaoZhao2026} also distinguishes operators that commute from operators that anticommute with $H$, which is exactly the distinction that will matter in Sec.~\ref{sec:algebra}. The theorem says which Chern numbers are \emph{allowed}; it never says that a Chern number \emph{is} nonzero once the symmetries are broken, and it does not apply at gap closings. We record it because its BdG specialization is enforced by \emph{two} operators.

\begin{proposition}[Glide lift]\label{prop:glide}
In the Nambu basis above let $\Gt=i\,\tau_z\otimes\sigma_x$ and $g(\ka)=(-k_x,k_y+\pi)$. For $p+ip$ pairing and any $\mu$, $\Gt H_{\rm BdG}(\ka)\Gt^\dagger=H_{\rm BdG}(g\ka)$ exactly. $\Gt$ is unitary and orientation-reversing, hence $C=0$ whenever the state is gapped.
\end{proposition}
The phase $i$ is a charge rotation of the particles by $\theta=\pi/2$, which acts as $e^{\pm i\theta}$ on the particle and hole blocks (relative phase $-1$): pairing carries charge two, so the glide may be accompanied by a global $U(1)$ rotation $e^{i\theta}$ of the particles, and the covariance condition on the pairing block is $\Delta(g\ka)=e^{2i\theta}\sigma_x\Delta(\ka)\sigma_x$. For $p+ip$ one finds $\theta=\pi/2$ ($e^{2i\theta}=-1$), which absorbs the sign of $\sin(k_y+\pi)$. The diagonal pairing block
\begin{equation}
\setlength{\arraycolsep}{1pt}
\Delta_{\rm d}(\ka)=\Dd\begin{pmatrix}0&e^{ik_y}-e^{-i(2k_x+k_y)}\\ e^{i(2k_x+k_y)}-e^{-ik_y}&0\end{pmatrix}
\end{equation}
produces under $g$ a harmonic $e^{i(2k_x-k_y)}$ absent from the original, so no phase $\theta$ can restore $\Gt$: a diagonal pairing of any complex phase breaks the glide lift.

\begin{proposition}[Hidden antiunitary obstruction]\label{prop:A}
Let $\cA=(\tau_z\otimes\sigma_x)\cK$ and $g'(\ka)=(k_x,k_y+\pi)$. Then $\cA H_{\rm BdG}(\ka)\cA^{-1}=H_{\rm BdG}(g'\ka)$ holds for $p+ip$ pairing, and with diagonal pairing if and only if $\Dd^*=-\Dd$, i.e.\ $\mathrm{Re}\,\Dd=0$. $\cA$ is antiunitary and orientation-preserving, hence $C=0$ whenever it survives and the state is gapped.
\end{proposition}

\begin{figure*}[!t]
\centering\includegraphics[width=0.96\textwidth]{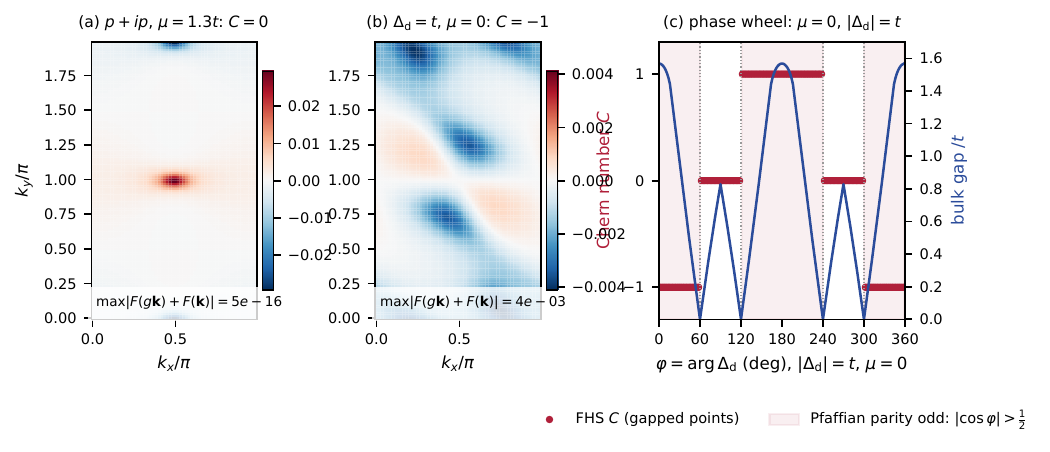}
\caption{Anatomy of Theorem~\ref{thm:confine}. (a) Berry curvature (FHS plaquette flux, $60\times120$) of $p+ip$ at $\mu=1.3t$: pointwise antisymmetric under the glide, $\max|F(g\ka)+F(\ka)|=5\times10^{-16}$, $C=0$. (b) With $\Dd=t$, $\mu=0$ the antisymmetry is broken ($4\times10^{-3}$) and $C=-1$. (c) Phase wheel at fixed $|\Dd|=t$, $\mu=0$: FHS Chern number at gapped points (red) and continuous-momentum gap (blue). The gap closes at $\varphi=60^\circ,120^\circ,240^\circ,300^\circ$; $C=\mp1$ for $|\cos\varphi|>\tfrac12$ (shaded, the Pfaffian-odd region of Prop.~\ref{prop:pf}) and $C=0$ for $|\cos\varphi|<\tfrac12$, including the range $60^\circ<\varphi<120^\circ$ where (except at $90^\circ$) $\mathrm{Re}\,\Dd\neq0$ but $C=0$.}
\label{fig:theorem}
\end{figure*}

Together, Props.~\ref{prop:glide} and \ref{prop:A} give a \emph{necessary} condition: $C\neq0$ requires $\mathrm{Re}\,\Dd\neq0$. The converse is false. A purely imaginary $\Dd$ breaks $\Gt$ but keeps $\cA$: at $\Dd=it$, $\mu=0$ the gap is $0.8284t$ and $C=0$. Rotating the phase $\varphi=\arg\Dd$ at fixed $|\Dd|=t$, $\mu=0$ (Fig.~\ref{fig:theorem}c), $C=-1$ for $|\varphi|<60^\circ$, the gap closes at $\varphi=60^\circ$, and $C=0$ for the whole open range $60^\circ<\varphi<120^\circ$, where (except at $\varphi=90^\circ$) $\mathrm{Re}\,\Dd\neq0$ and both obstructions are broken but the Chern number vanishes nonetheless (e.g.\ $\varphi=88^\circ$: gap $0.7786t$, $C=0$). Where the Chern number \emph{is} odd is decided by a different, standard criterion.

\begin{proposition}[Pfaffian parity]\label{prop:pf}
At $\Dz=t$, write $m=\mu/t$, $d=\Dd/t$. At the four particle--hole-invariant momenta $\mathbf K\in\{(0,0),(0,\pi),(\tfrac\pi2,0),(\tfrac\pi2,\pi)\}$ the matrix $B(\mathbf K)=H_{\rm BdG}(\mathbf K)\tau_x$ is antisymmetric, with Pfaffians (in units of $t^2$)
\begin{equation}
\begin{gathered}
8-m^2,\quad 8-m^2,\\
4-m^2-4|1+d|^2,\quad 4-m^2-4|1-d|^2 .
\end{gathered}
\end{equation}
For a gapped state, $(-1)^C=\prod_{\mathbf K}\mathrm{sgn}\,\mathrm{Pf}\,B(\mathbf K)$, i.e.
\begin{equation}
C\ \text{odd}\iff\big[m^2+4|1+d|^2-4\big]\big[m^2+4|1-d|^2-4\big]<0 .
\label{eq:parity}
\end{equation}
\end{proposition}
The parity relation is the class-D statement that the Chern parity equals the product of the one-dimensional class-D invariants of the two particle--hole-invariant lines $k_y=0,\pi$, each given by the Pfaffian signs at its two invariant momenta \cite{Sato2010,Kitaev2001}; the four Pfaffians of this model are verified here symbolically and numerically (the numerical Pfaffians agree with the formula at every point of the $81\times81$ grid, Appendix~\ref{app:methods}). For real $d$ Eq.~\eqref{eq:parity} reads $m^2+4(|d|-1)^2<4$: the odd-$C$ phases are two ellipses centered at $(\mu,\Dd)=(0,\pm t)$, touching the Klein-bottle line $\Dd=0$ only at the nodal point $\mu=\Dd=0$. On the phase wheel it reads $|\cos\varphi|>\tfrac12$, which reproduces the $60^\circ$ boundaries of Fig.~\ref{fig:theorem}c. The criterion gives the parity of $C$, not its sign or its value, and it presupposes a gap.

\begin{figure*}[!t]
\centering\includegraphics[width=0.96\textwidth]{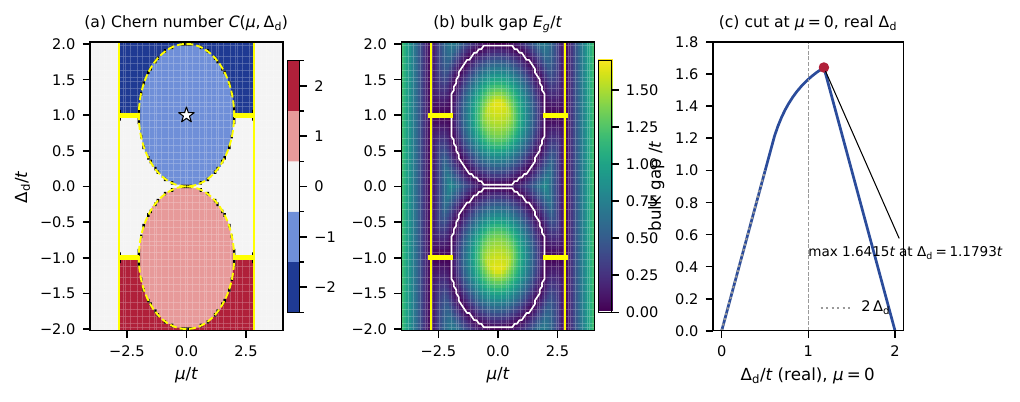}
\caption{Pairing phase diagram at $\Dz=t$ on an $81\times81$ grid of $(\mu,\Dd)\in[-4,4]\times[-2,2]$, real $\Dd$. (a) FHS Chern number ($72\times144$ momenta) at the $6422$ grid points whose continuous-momentum gap exceeds the threshold $0.02t$; black: below the threshold ($139$ points). The gap closings of the section are drawn analytically in yellow: the Pfaffian ellipses $\mu^2+4(t\pm\Dd)^2=4t^2$ (dashed), the lines $\mu=\pm2\sqrt2\,t$ (solid), and the nodal segments $\Dd=\pm\Dz$, $2t\le|\mu|\le2\sqrt2\,t$ (thick). Star: benchmark $(0,t)$. The FHS integer parity agrees with Eq.~\eqref{eq:parity} at all $6422$ points. (b) Continuous-momentum bulk gap; white contour: Pfaffian-odd region; yellow as in (a). (c) Cut at $\mu=0$: the gap opens as $2|\Dd|$ for small real $\Dd$ ($\Dd=0.05t$ gives $0.1000t$), passes $1.5686t$ at the benchmark and reaches $1.6415t$ at $\Dd=1.1793t$ on this line.}
\label{fig:phase}
\end{figure*}

Figure~\ref{fig:phase} shows the phase diagram. Every parameter point is treated in three steps (Appendix~\ref{app:methods}): the gap is minimized over continuous momentum (a dense grid, shrinking local searches from several starting points, the four particle--hole-invariant momenta evaluated exactly, and a fine scan of the line $k_y=\pi$), the Pfaffian parity is evaluated analytically, and the FHS Chern number \cite{FukuiHatsugaiSuzuki2005} is computed only where the gap exceeds the threshold $0.02t$; it is an integer to machine precision at every such point and its parity agrees with Eq.~\eqref{eq:parity} everywhere. The count of $6422$ gapped points is a statement about grid points above a threshold, not a global gap certificate; the gap closings that we identify are found analytically and form three families (a $161\times161$ scan of the section, with every low-gap grid point refined by the multi-start minimizer, finds no gap closing away from them, Appendix~\ref{app:methods}, but we do not claim exhaustiveness). (A) At $\mathbf K=(0,0),(0,\pi)$ the Pfaffian $8-m^2$ vanishes for $\mu=\pm2\sqrt2\,t$ \emph{for every real $\Dd$}: the whole lines $\mu=\pm2\sqrt2\,t$ are nodal, and they are the boundaries of the $C=\mp2$ phases. Since $2\sqrt2$ is not a grid value, a grid alone does not see this family ($\mu=2.8t$ gives $0.028t$, above the threshold). (B) The Pfaffian ellipses $\mu^2+4(t\pm\Dd)^2=4t^2$, on which the energy at $(\pi/2,\pi)$ or $(\pi/2,0)$ vanishes; these are the odd/even boundaries. (C) On the line $\Dd=\Dz$ the pairing matrix vanishes identically at $k_y=\pi$, so the BdG energy vanishes wherever the normal-state band crosses $\mu$, i.e.\ at $|\mu|=2t\sqrt{1+\cos^2k_x}$, which has solutions for $2t\le|\mu|\le2\sqrt2\,t$: a nodal segment (numerically $|E|<6\times10^{-16}t$ at the analytic node for $\mu=2.0,2.1,\dots,2.828$), which a coarse momentum grid misses, returning spurious gaps of order $0.03$--$0.05t$ and a spurious Chern integer. Consequently the line $\Dd=0$ is gapped with $C=0$ except at the node $\mu=\Dd=0$ and at $\mu=\pm2\sqrt2\,t$ (gaps $0.0616t$, $0.3854t$, $0.8284t$ at $\mu=0.5t,1.3t,2t$; $1.1716t$ at $4t$); it is not a critical line. Two odd-$C$ phases, $C=-1$ ($\Dd>0$) and $C=+1$ ($\Dd<0$), fill the ellipses, which end at $|\mu|=2t$ on the line $\Dd=\pm\Dz$; $\Dd\to-\Dd$ maps $C\to-C$. Gapped even phases $C=\mp2$ occupy $|\Dd|>\Dz$, $|\mu|<2\sqrt2\,t$ outside the ellipses, with gaps up to $0.696t$ (at $\mu=\pm1.9t$, $\Dd=-2t$; $\mu=2t$, $\Dd=2t$ gives $C=-2$ with gap $0.686t$). At the benchmark $(\mu,\Dd)=(0,t)$, $C=-1$ with bulk gap $1.5686t$. On the line $\mu=0$ the gap is maximal at $\Dd=1.1793t$ with $E_g=1.6415t$, the point where a branch of the minimum at a particle--hole-invariant momentum crosses a branch at a generic momentum (a one-dimensional search; the grid maximum $1.6313t$ at $\Dd=1.15t$ is a grid value, and we do not claim a certified global optimum of the full diagram). For complex $\Dd$ the section changes; the phase wheel of Fig.~\ref{fig:theorem}(c) is the only complex cut we show. The Chern numbers were cross-checked at ten parameter points, including $C=\pm1$, $C=-2$ and $C=0$ points, against the real-space Bott index on $16\times16$ and $24\times24$ tori (Appendix~\ref{app:methods}, Table~\ref{tab:bott}).

\section{Symmetry algebra}\label{sec:algebra}

The operators met so far close into a small projective algebra, which we now state with the distinction that Ref.~\cite{XiaoZhao2026} rightly insists on: an element $S$ with $SH(\ka)S^{-1}=+H(g\ka)$ is a symmetry, one with $SH(\ka)S^{-1}=-H(g\ka)$ is a spectral antisymmetry, which maps negative-energy states to positive-energy ones and is not a symmetry of a fixed Hamiltonian. Particle--hole conjugation $\cC$ is the prime example.

The glide lift squares to fermion parity, $\Gt^2=-\mathbb 1=(-1)^F$: pairing promotes the $\mathbb Z_2$ glide to a $\mathbb Z_4$ projective representation. Further, $\cA^2=+1$, $\cC^2=+1$, $[\cC,\Gt]=0$, and $\cA=(-1)^F R\,\Gt$ with $R=-i\cK\circ(k_x\to-k_x)$ the normal-state real structure: the antiunitary obstruction is not an independent postulate but the composite of the glide with the real structure. Closing the set $\{\Gt,\cA,\cC,R\}$ under products as (matrix, antiunitary flag, affine $\ka$-map) triples gives sixteen elements---eight unitary and eight antiunitary, all with scalar factors $\pm1$ only---which reduce to eight distinct $\ka$-maps modulo $\pm1$. Its unitary part satisfies the quaternion relations,
\begin{align}
i&=\Gt,\quad j=\cC\cA=-i\tau_y\otimes\sigma_x,\quad k=ij=-i\tau_x,\nonumber\\
i^2&=j^2=k^2=ijk=-1=(-1)^F ,
\end{align}
with $-1$ of the quaternions the fermion parity; as an abstract group generated by $i$ and $j$ this is $Q_8$. But the relations are relations of an \emph{operator algebra containing spectral antisymmetries}: since $\cC$ anticommutes with $H$ in the sense above, $j=\cC\cA$ and $k=\Gt j$ satisfy $jH(\ka)j^{-1}=-H(-k_x,-k_y+\pi)$, $kH(\ka)k^{-1}=-H(k_x,-k_y)$ (the affine maps of Table~\ref{tab:survival}; numerically exact), and only $\{\pm1,\pm\Gt\}$ among the eight unitary elements are symmetries of the $p+ip$ Hamiltonian. Table~\ref{tab:survival} lists the sixteen elements by type; the list is the closure of the four generators and says nothing about symmetries of other forms. Likewise the charge $U(1)$ is not a symmetry of a fixed pairing background: $e^{i\theta\tau_z}H(\Delta)e^{-i\theta\tau_z}=H(e^{2i\theta}\Delta)$ (residual $6\times10^{-16}$ against the rotated Hamiltonian, $1.6$ against the fixed one). The finite spectral-symmetry algebra has three squares $\Gt^2$, $\cD^2$ (Sec.~\ref{sec:KB}) and $j^2$ equal to $(-1)^F$; the continuous charge rotations act on the pairing family rather than belonging to this finite algebra.

\begin{table}[!t]
\caption{The sixteen-element closure of $\{\Gt,\cA,\cC,R\}$ for $p+ip$ pairing ($\mu=1.3t$), and which elements survive the diagonal pairing. ``sym'' means $SH(\ka)S^{-1}=H(g\ka)$, ``anti'' means $SH(\ka)S^{-1}=-H(g\ka)$ (residuals $<10^{-15}$); ``--'' means neither. Elements come in pairs $\pm S$.}
\label{tab:survival}
\centering\scriptsize\setlength{\tabcolsep}{3pt}
\begin{tabular}{lcccccc}
\toprule
$\pm S$ & type & $S^2$ & $g$ & $p{+}ip$ & $\Dd{=}it$ & $\Dd{=}t$\\
\midrule
$1$ & unit. & $+1$ & id & sym & sym & sym\\
$\Gt$ & unit. & $-1$ & $(-k_x,k_y+\pi)$ & sym & -- & --\\
$\cC\cA=j$ & unit. & $-1$ & $(-k_x,-k_y+\pi)$ & anti & anti & --\\
$\Gt j=k$ & unit. & $-1$ & $(k_x,-k_y)$ & anti & -- & --\\
$\cA$ & antiunit. & $+1$ & $(k_x,k_y+\pi)$ & sym & sym & --\\
$R$ & antiunit. & $+1$ & $(-k_x,k_y)$ & sym & -- & --\\
$\cC$ & antiunit. & $+1$ & $(-k_x,-k_y)$ & anti & anti & anti\\
$\cC\Gt$ & antiunit. & $-1$ & $(k_x,-k_y+\pi)$ & anti & -- & --\\
\bottomrule
\end{tabular}
\end{table}

A census of the sixteen elements shows that the unitary elements squaring to $-1$ other than $\pm\Gt$ are antisymmetries, while the antiunitary elements squaring to $+1$ other than $\pm\cC$ are symmetries---the quaternionic/real dichotomy is locked to the symmetry/antisymmetry distinction rather than to unitarity alone. Real $\Dd$ collapses the symmetry part to $\{\pm1\}$ and the antisymmetry part to $\{\pm\cC\}$: at $C=-1$ the BdG Hamiltonian has no unitary orientation-reversing and no antiunitary orientation-preserving symmetry, as Theorem~\ref{thm:confine} requires, and none of the mirror-type elements of the closure (unitary or antiunitary) survives, because the diagonal pairing singles out one diagonal; whether some operation outside this closure survives is not decided by the table. We will meet the consequence of this on the real-space Klein bottle.

\section{The real-space Klein bottle}\label{sec:KB}

We now take the quotient of the real-space lattice, Fig.~\ref{fig:lattice}(b),
\begin{equation}
(x,\,y+L_y)\ \sim\ \big((1-x)\bmod L_x,\ y\big),\qquad L_x\ \text{even},
\label{eq:quotient}
\end{equation}
a bond-centered mirror composed with a half-period translation (a ``reflection twist'' in the language of Ref.~\cite{SeibergZhang2026}, whose consistency condition $L_1(L_2-1)$ even is satisfied for even $L_x$ and any $L_y$; all sizes used below are even). Two constructions are used: (a) an explicit quotient lattice in which every bond leaving the top row lands at the mirrored column of the bottom row with the same amplitude it would have in the bulk (the ``literal'' gluing, with the Landau-gauge sign taken from the source column), and (b) the covering torus $L_x\times2L_y$ with the deck transformation $\mathcal D=\mathcal G_{(-1)^y}M_{x\to1-x}T_{L_y\hat y}$. In the normal state $\mathcal D^2=+1$, the two constructions agree to $10^{-14}$ (on $12\times12$, with and without the dimerization of Appendix~\ref{app:Z2}), and the $\nu=1$ Klein-bottle insulator of Appendix~\ref{app:Z2} ($\lambda=-0.6t$) is fully gapped on the closed surface (lowest $|E|=0.6t$, the bulk gap, on $12\times12$ and $24\times24$; no in-gap state), as it must be, since its edge states need an edge. A useful property of the quotient is that the $x$-translations do not survive it: $T_a$ conjugates the gluing map $x\to1-x$ into $x\to1-x+2a$, so only $T_{L_x/2}$ ($2a\equiv0\bmod L_x$) is compatible with the quotient geometry, and the seam has no exact momentum quantum number. Compatibility with the geometry is not yet a symmetry of the Hamiltonian: $T_{L_x/2}$ shifts the Landau-gauge pattern $(-1)^x$ by $L_x/2$ columns, so it commutes with $H$ only for $L_x\equiv0\pmod4$ (residual $0$ for $L_x=8,12,20$; residual $2$ for $L_x=6,10,22$).

\paragraph{Glide-covariant pairing: deck eigenblocks, not two quotients.} For $p+ip$ ($C=0$, $\mu=1.3t$) the deck transformation lifts to BdG only with the $U(1)_{\pi/2}$ phase of Prop.~\ref{prop:glide}, $\cD=\mathrm{diag}(i\mathcal D,-i\mathcal D^*)$, so that $[H_{\rm cover},\cD]=0$ and
\begin{equation}
\cD^2=-1=(-1)^F ,
\end{equation}
the lattice avatar of the statement that fermions on a non-orientable manifold require a Pin structure \cite{Witten2016RMP}. $H_{\rm cover}$ therefore splits into the eigenblocks $\cD=\pm i$; both are gapped with equal gaps $0.3854t$ (equal to the cover gap) and identical spectra (maximal difference $2\times10^{-14}$), and the block energy is exactly half the cover energy. The two blocks are not independent Pin-sector quotients: $\cC$ is antilinear, so $\cD\psi=i\psi$ implies $\cD(\cC\psi)=-i\,\cC\psi$, i.e.\ particle--hole conjugation \emph{exchanges} the two blocks, and a single block is not closed under the Nambu real structure (its standard particle--hole residual is $2.0$, the maximal value). Each block is a spectral block of the covering Hamiltonian, not a BdG Hamiltonian of fermions living on the quotient with their own conjugates. Nor is the literal gluing (a) equivalent to a block: for the sixteen literal gluings in which the cross-seam hopping and pairing carry $c$-number factors $(\eta_h,\eta_p)\in\{\pm1,\pm i\}^2$, all sixteen are Hermitian, they fall into seven isospectral classes with gaps between $0.2946t$ and $0.5575t$ (on $12\times12$), and none reproduces the block spectrum (minimal deviation $0.156t$). The classification of $p+ip$ Hamiltonians on a lattice Klein bottle---which gluings are canonical, and how a Pin structure is to be imposed on the quotient fermions rather than read off from the cover---is therefore an open problem that we do not settle here; the covariant state suffers no frustration on the cover, and that is all we use.

\paragraph{Chiral pairing: the seam is a chirality wall.} For $C=-1$ ($\Dz=\Dd=t$, $\mu=0$) there is no deck lift of the uniform state: a unitary, energy-preserving, orientation-reversing lift would contradict Theorem~\ref{thm:confine}, and Sec.~\ref{sec:algebra} showed that no orientation-reversing element of the sixteen-element closure survives at $C=-1$. (A cover that already contains a chirality wall can of course be deck-covariant; the obstruction is to a lift of the \emph{uniform} chiral state.) Any gluing that keeps the uniform gapped $C=-1$ bulk on both sides of the seam, reconstructs the state only within a finite width of the seam, and does not close the bulk gap therefore produces a chirality domain wall, $C=-1\,|\,C=+1$ with $\Delta C=2$, along the seam. (The obstruction by itself does not forbid other resolutions, such as a bulk gap closing or a vanishing order parameter; it is the fixed pairing background of this paper that selects the wall.) A wall of this type is known to carry two co-propagating chiral Majorana channels \cite{Serban2010}; the flux-free Majorana modes forced by a M\"obius geometry were found in Ref.~\cite{Quelle2016}. What the closed lattice adds is the possibility to look at the seam with no other defect on the surface.

\begin{figure*}[!t]
\centering\includegraphics[width=0.98\textwidth]{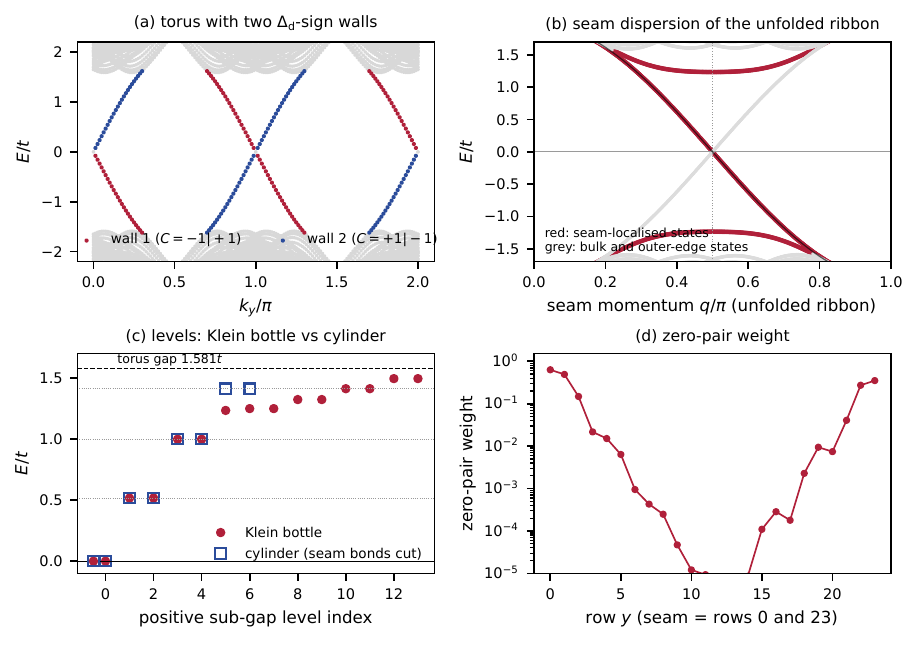}
\caption{The seam of the chiral state ($\Dz=\Dd=t$, $\mu=0$). (a) Reference: a $64$-column torus with two hand-made walls $\Dd\to-\Dd$ (mixed real-$x$/Bloch-$k_y$ representation). Each wall carries \emph{two} chiral Majorana branches with the \emph{same} velocity, crossing zero at $k_y=0$ and $k_y=\pi$ (red: wall 1, slopes $-1.998t$ and $-1.998t$; blue: wall 2, $+1.998t$ twice). (b) Seam dispersion of the unfolded ribbon: the Klein bottle is cut between two bulk rows far from the seam and its upper half reflected, $x\to1-x$, which makes the strip exactly $T_2$-invariant; its Bloch reduction (site momentum $q$, cell phase $e^{2iqm}$, $48$ rows) gives the spectrum shown. Red: seam-localised states (weight $>0.5$ within three rows of the seam): two branches of one sign of energy on each side of $q=\pi/2$, degenerate to $2.5\times10^{-9}t$ outside a window $|q-\pi/2|<0.011$ in which they hybridise with the outer-edge crossings at finite $L_y$, following $E=2t\cos q$ (black) with velocity $-2.0000t$ at the crossing, and a seam-bound band near $\pm1.24t$. Grey: bulk states and the ordinary chiral edge branches of the strip's two outer edges (opposite velocity, far from the seam). (c) Positive sub-gap levels of the Klein bottle (red) and of the cylinder obtained by setting every cross-seam bond to zero (blue squares); dotted: $2t\sin(2\pi n/L_x)$; dashed: the $24\times24$ torus gap $1.5813t$. The low-energy content is the same; the Klein bottle has additional seam-bound levels between $1.24t$ and $1.50t$. (d) Row profile of the near-zero pair: $87\%$ of its weight within two rows of the seam (rows $0,1,22,23$), uniform along the seam row to $10^{-14}$.}
\label{fig:seam}
\end{figure*}

Figure~\ref{fig:seam} shows the result on $24\times24$. The bulk keeps the torus gap ($1.5813t$ at this size; $1.5686t$ in the thermodynamic limit). On the seam appears an in-gap tower
\begin{equation}
E_n=2t\,\sin\!\Big(\frac{2\pi n}{L_x}\Big),
\label{eq:tower}
\end{equation}
whose first level is reproduced to better than $10^{-8}t$ for $L_x=16,20,24,28,32$ (at $L_y=20$: $0.76536687$, $0.61803399$, $0.51763809$, $0.44504187$, $0.39018065$), and whose second level is reproduced for $L_x\ge20$ (for $L_x=16$ an additional seam-bound level at $1.236t$ lies below $2t\sin(4\pi/16)=1.414t$). Each positive level $E_n$, $n\ge1$, is doubly degenerate, and the count of states below $1.2t$ is $6,10,14$ for $L_x=16,24,32$.

What these numbers mean is settled by three comparisons. First, the reference torus with two hand-made walls, Fig.~\ref{fig:seam}(a): each $\Delta C=2$ wall carries two chiral Majorana branches with the same velocity $2t$, whose zero crossings sit at the wall momenta $k_y=0$ and $\pi$, displaced by $\pi$. Second, the seam itself. The closed surface has no translation along the seam, so seam momentum is not a quantum number there: a Fourier analysis of the closed-system eigenstates gives spectral weights only (the $E_1$ doublet of $24\times24$ has $66\%$ of its weight on a negative-slope arm and $34\%$ on the opposite arm, with the site momenta $p$ and $p+\pi$ equivalent by the two-site cell), and cannot by itself count channels. A translation-invariant description exists after unfolding: cutting the bonds between two bulk rows far from the seam and reflecting the upper half, $x\to1-x$, turns every cross-seam bond into an ordinary bond of a strip that is exactly invariant under $T_2$ (residual $0$). Its Bloch reduction, Fig.~\ref{fig:seam}(b), shows the seam content directly. Tracking seam-localised in-gap states ($|E|<1.1t$, weight $>0.5$ within three rows of the seam) by sign of energy, for $0.32\pi\le q\le0.68\pi$ excluding a window $|q-\pi/2|<0.011$, one finds at every $q<\pi/2$ exactly two seam states of positive energy and none of negative energy, and at every $q>\pi/2$ the reverse; the two states of equal sign are degenerate to $2.5\times10^{-9}t$. The branch is $E(q)=2t\cos q$ to the same precision (energies $\pm0.0399973t$ at $q=\pi/2\mp0.02$, equal to $2t\sin0.02$), i.e.\ velocity $-2.0000t$ at the crossing $q=\pi/2$. In the excluded window the seam pair is degenerate with the zero crossings of the strip's two outer edges, and at finite $L_y$ the four states hybridise: at $q=\pi/2$ the near-zero energies are $\pm3.0,\pm1.4\times10^{-4}t$, $\pm5.7,\pm4.0\times10^{-6}t$, $\pm5.5,\pm2.3\times10^{-8}t$ and $<10^{-9}t$ for $L_y=32,48,64,96$, an exponentially decaying seam--edge coupling. A seam-bound band near $\pm1.24t$ and the outer-edge branches (opposite velocity, far from the seam) are also visible. Both seam branches have the same sign of velocity: the seam carries two \emph{co-propagating} chiral Majorana branches, exactly the content of a $\Delta C=2$ wall, and not a right-mover and a left-mover. Two co-propagating Majorana branches are one chiral complex (Dirac) fermion, $c=1$. The degeneracy of the two branches at every $q$ is an observed property of this seam; in the closed system it is why every positive level appears twice (particle--hole conjugation maps $(q,E)$ to $(-q,-E)$ and does not by itself produce it). Three momenta should not be confused: the Dirac points $(\pi/2,\pi/2)$, $(\pi/2,3\pi/2)$ of the normal state; the crossings $k_y=0,\pi$ of the two-wall reference, which are the Fermi points of the wall branches along $y$; and the seam momentum $q=\pi/2$ of the unfolded ribbon, the Fermi point of the wall branches along $x$ modulo the two-site cell. Third, the cylinder, Fig.~\ref{fig:seam}(c): setting every cross-seam bond to zero leaves two open edges, each with one chiral Majorana branch of the $C=-1$ (or, in the mirrored frame, $C=+1$) bulk, and the low-energy count and the first levels are \emph{identical} to those of the Klein bottle ($2$ near-zero states and $6/10/14$ states below $1.2t$ for $L_x=16/24/32$, $E_1$ agreeing to $4\times10^{-11}t$). The non-orientable gluing does not create the gapless content: it reconnects the two edge channels of a cylinder into the two channels of one wall, at the same energies, and adds a family of seam-bound levels between $1.24t$ and $1.50t$ which the cylinder lacks. The level counts alone cannot discriminate a single $c=\tfrac12$ ring of length $2L_x$ from two channels; the unfolded ribbon resolves the seam content as two co-propagating branches.

\paragraph{The near-zero pair.} For periodic boundary conditions along $x$ and $L_x\equiv0\pmod 4$ the tower has a level at $n=0$: a pair of BdG states at $\pm\epsilon$ with $\epsilon=2.4\times10^{-3}t$, $1.2\times10^{-4}t$, $8.4\times10^{-6}t$, $3.0\times10^{-6}t$, $3.4\times10^{-8}t$ for $L_y=12,16,20,24,32$ ($L_x=24$), i.e.\ exponentially small in the distance between the two sides of the seam. It is one complex fermion mode---the mode of the seam Dirac fermion at its Fermi point---whose two Majorana components live on the same seam, with $87\%$ of their weight within two rows of it and a density uniform along it (Fig.~\ref{fig:seam}d). There is no vortex anywhere. Three facts limit what this pair is:

\emph{Circumference and holonomy.} The pair is a momentum-quantization effect. The gluing leaves a choice of flat $\mathbb Z_2$ holonomy $\theta\in\{0,1\}$ along the non-contractible $x$ cycle (the antiperiodic case is a sign on one column of $x$ bonds, with the cross-seam signs adjusted so that every plaquette keeps its $\pi$ flux), and the seam levels are
\begin{equation}
\begin{gathered}
E_n=2t\cos q_n,\quad q_n=\frac{(2n+\theta)\pi}{L_x},\\
n=0,\dots,\tfrac{L_x}{2}-1,
\end{gathered}
\label{eq:tower2}
\end{equation}
the signed seam branch of Fig.~\ref{fig:seam}(b) sampled at the allowed seam momenta; each level comes with its particle--hole partner $-E_n$ at $-q_n$, and every level is doubly degenerate (the two branches). For $\theta=0$ and $L_x\equiv0\pmod4$ the positive members reproduce Eq.~\eqref{eq:tower}. The ideal branch in Eq.~\eqref{eq:tower2} has a zero level iff the Fermi momentum $\pi/2$ is an allowed $q_n$; the finite closed system retains the splitting $\epsilon$ reported above. The condition is that $L_x/2+\theta$ is even: $L_x\equiv0\pmod4$ in the periodic sector and $L_x\equiv2\pmod4$ in the antiperiodic one. Numerically ($L_y=24$): with periodic holonomy $L_x=8,12,16,20$ have the $3\times10^{-6}t$ pair while $L_x=6,10,14,18,22$ have lowest level $2t\sin(\pi/L_x)=1.000,0.618,0.445,0.347,0.285t$ and no near-zero state; with antiperiodic holonomy the two groups exchange roles exactly (e.g.\ on $24\times24$ the lowest levels become $0.261052t=2t\sin(\pi/24)$, four-fold, and $0.765367t=2t\sin(3\pi/24)$). The dependence is not specific to the gluing: the cylinder obtained by cutting every cross-seam bond admits the same flat twist along its non-contractible cycle, and with it the same exchange of sectors: for $L_x=6,\dots,22$ the lowest level of each sector coincides with that of the Klein bottle---a near-zero pair in one sector ($3\times10^{-6}t$ and $4\times10^{-6}t$ on the two surfaces) and the four-fold level $2t\sin(\pi/L_x)$ in the other, agreeing to $10^{-9}t$. The near-zero pair is a property of one sector, not of the Klein bottle as such.

\begin{figure}[!t]
\centering\includegraphics[width=\columnwidth]{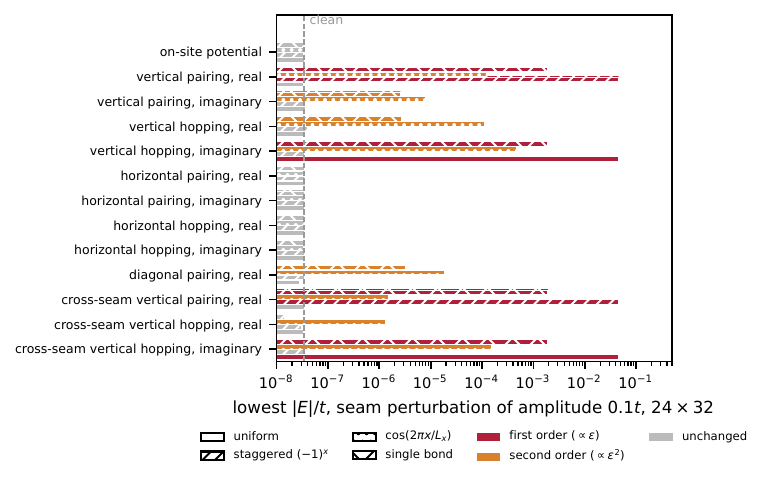}
\caption{Which seam perturbations split the near-zero pair ($24\times32$; bars: lowest $|E|$ at amplitude $0.1t$; clean value $3.4\times10^{-8}t$, dashed). Each of the thirteen term families is tested uniform along the seam, with a $(-1)^x$ modulation, with a $\cos(2\pi x/L_x)$ modulation, and on a single bond (hatching); the colour gives the order from the amplitude scaling of the splitting over $\epsilon=0.05$--$0.2t$: red first order ($\propto\epsilon$), orange second order ($\propto\epsilon^2$), grey unchanged (below $3\times10^{-7}t$ at all amplitudes).}
\label{fig:pert}
\end{figure}

\emph{Local perturbations.} Particle--hole symmetry only requires the spectrum to come in $\pm E$ pairs; two Majorana components on the same seam may be coupled by $i\epsilon\gamma_1\gamma_2$ by any local term that is not forbidden by a further symmetry. We tested thirteen families of bond and site terms on the seam rows, each uniform along the seam, with a $(-1)^x$ modulation, with a $\cos(2\pi x/L_x)$ modulation, and on a single bond---$52$ entries---at amplitudes $\epsilon=0.025,0.05,0.1,0.2\,t$ on $24\times32$, where the clean pair ($3.4\times10^{-8}t$) is small enough not to mask weak effects (on $24\times16$ the clean value $1.2\times10^{-4}t$ hides them). The order of each entry is the log--log slope of the splitting against $\epsilon$, not a threshold at one amplitude (Fig.~\ref{fig:pert}). \emph{First order} (slope $1.00$, eight entries): a real pairing on the vertical seam bonds with a $(-1)^x$ modulation, $\Delta_y(x,0)=it+\epsilon(-1)^x$, and a uniform \emph{imaginary} hopping on the same bonds, the cross-seam versions of both, and a single bond of any of these four types. The coherent sums split the pair to $0.459\,\epsilon$ (i.e.\ $0.0459t$ at $\epsilon=0.1t$), with only weak variation over the tested widths ($0.0483,0.0457,0.0459,0.0459t$ for $L_y=12,16,24,32$), and split the first tower level into $0.476t$ and $0.569t$; a single bond gives $0.0191\,\epsilon$, which is $1/L_x$ of the coherent value, as expected for one term of the sum; the same staggered pairing on a bulk row produces a much weaker response in that scan, with no comparable lifting of the pair. This finite comparison is not exact invariance under bulk perturbations. \emph{Second order} (slope $1.96$--$2.01$, eleven entries): the $\cos(2\pi x/L_x)$-modulated vertical terms, which carry seam momentum $2\pi/L_x$ and connect $q=\pi/2$ to neighbouring momenta---real pairing, real hopping, imaginary hopping and imaginary pairing, with $\epsilon^2\times(0.012,0.011,0.047,0.0008)$, the cross-seam versions of the first three ($\epsilon^2\times(1.5\times10^{-4},1.3\times10^{-4},0.015)$), and the diagonal pairing ($0.0019\,\epsilon^2$)---and single bonds of real hopping, imaginary pairing and diagonal pairing ($\epsilon^2\times(2.7,2.5,3.2)\times10^{-4}$). \emph{Unchanged} ($33$ entries, below $3\times10^{-7}t$ at all amplitudes): every on-site potential ($4$), every horizontal term ($16$), and among the vertical and diagonal terms the uniform real pairing, the uniform and staggered imaginary pairings, the uniform and staggered real hoppings, the staggered imaginary hopping, the uniform and staggered diagonal pairings, and the cross-seam analogues (uniform real pairing; uniform, staggered and single-bond real hopping; staggered imaginary hopping). The first-order class is explained by the unfolded ribbon: a $T_2$-invariant seam term (uniform or $(-1)^x$-modulated) acts within the $q=\pi/2$ sector, and its first-order effect is the $2\times2$ matrix in the seam-projected basis of the near-zero subspace there (at finite $L_y$ this subspace is four-dimensional, seam pair plus outer-edge crossings, and the basis is the pair of combinations with maximal seam weight); that matrix has eigenvalues $\pm0.04588t$ at $\epsilon=0.1t$ for exactly the four coherent first-order families and vanishes ($<10^{-16}$) for all other $T_2$-invariant entries, in agreement with the closed-system value $0.0459t$. This is a numerical selection rule of the specific seam wavefunctions at the sizes studied, not a symmetry protection. Random on-site disorder of strength $W=2t$ on all sites (three realizations, $24\times16$) lifts the pair to $(2$--$7)\times10^{-3}t$ with the next level at $0.48t$, a higher-order effect at that amplitude. The $0.1t$ counterexamples suffice to rule out protection against generic allowed local perturbations.

\emph{Symmetry.} We searched for lattice symmetries of the chiral Klein-bottle Hamiltonian among all compositions of $x$-translations, the mirrors $x\to1-x$ and $y\to L_y-1-y$, thirteen gauge patterns ($(-1)^x$, $(-1)^y$, $i^x$, $(-1)^{xy}$, \dots), four Nambu phases, particle--hole swap and complex conjugation ($6656$ candidates on $8\times6$): within this family the only symmetries are the identity and $T_{L_x/2}$ (the latter for $L_x\equiv0\pmod4$), and the only antisymmetries their compositions with $\cC$. The search excludes the listed templates only, not every conceivable lattice or canonical operation; the counterexamples above are what rule out protection, and we make no protection claim. The smallness of the clean pair's splitting is a momentum-quantization property of the clean gluing in the appropriate sector (Eq.~\eqref{eq:tower2}), which we describe but do not explain by a symmetry; the clean pair is a finite-size near-zero pair, not an exact zero mode.

\input{odd_results}

\section{Cost and correlations of the seam}\label{sec:cost}

\begin{figure}[!t]
\centering\includegraphics[width=0.9\columnwidth]{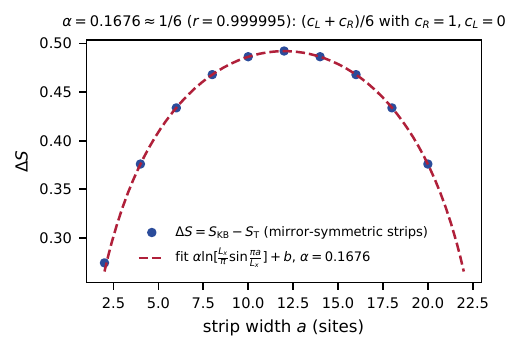}
\caption{Entanglement of mirror-symmetric strips on $24\times16$: $\Delta S=S_{\rm KB}-S_{\rm T}$ against $\ln[\tfrac{L_x}{\pi}\sin\tfrac{\pi a}{L_x}]$ has slope $\alpha=0.16765$ ($r=0.999995$; nine strips, five independent widths). $(c_L+c_R)/6=1/6$ for a chiral $c_R=1$ seam; a non-chiral $c=\tfrac12$ theory gives the same number, so the slope is consistent with, but does not by itself identify, the chiral Dirac seam of Sec.~\ref{sec:KB}.}
\label{fig:entangle}
\end{figure}

\paragraph{Entanglement.} An $x$-strip $[0,\ell)$ is not a geometrically local region on the Klein bottle, because the gluing identifies $x$ with $1-x$; every cross-seam mirror bond of an asymmetric strip is cut, producing an additional area-law contribution of $\approx0.44$ per unmirrored column (entropy, dimensionless; on $24\times16$, width $12$)---itself a measurable imprint of the non-orientable geometry. Any small region away from the seam is of course local; for a full-height strip, the choice symmetric about the mirror-fixed points ($x=\tfrac12$ bond centers) is the one that cuts no additional seam bonds, so that the subtraction of the torus reference isolates the seam. With such cuts, the Klein-bottle-minus-torus entropy difference $\Delta S(a)$ of a strip of width $a$ fits $\alpha\ln[\tfrac{L_x}{\pi}\sin\tfrac{\pi a}{L_x}]$ with
\begin{align}
\alpha&=0.16765\quad(r=0.999995),\nonumber\\
\tfrac{c_L+c_R}{6}\Big|_{c_R=1,c_L=0}&=0.16667 ,
\end{align}
Fig.~\ref{fig:entangle} \cite{CalabreseCardy2004,Peschel2003}. The single-interval coefficient $(c_L+c_R)/6$ of a chiral $c_R=1$ seam and $c/3$ of a non-chiral $c=\tfrac12$ theory coincide, so the slope is consistent with the chiral Dirac seam but does not identify it; the identification rests on Fig.~\ref{fig:seam}. The fit uses one size and five independent widths and is not a finite-size-scaled determination. The entanglement spectrum of a symmetric strip has a doubly degenerate minimal $\nu=\{0.1305,0.1305\}$, against a single exact $\nu=0$ for the torus reference. A four-row band containing the seam and a four-row band in the bulk have mutual information $0.039$ on the Klein bottle against $0.032$ for the same rows on the torus: the seam adds little long-range correlation to the bulk.

\paragraph{Seam tension.} The energetic price of the wall is a seam tension per seam site, which we define with two references, the torus of the same size and half the covering torus $L_x\times2L_y$:
\begin{align}
\tau_{\rm same}&=\big[E_{\rm KB}(L_x,L_y)-E_{\rm T}(L_x,L_y)\big]/L_x,\nonumber\\
\tau_{\rm cover}&=\big[E_{\rm KB}(L_x,L_y)-\tfrac12E_{\rm T}(L_x,2L_y)\big]/L_x .
\end{align}
For $C=-1$ at circumference $L_x=24$ they are $0.15834/0.15955$, $0.16004/0.15999$, $0.16003/0.16004$, $0.16004/0.16004\,t$ at $L_y=8,12,16,20$: the two references converge to the same value,
\begin{equation}
\tau(C{=}-1;L_x{=}24)=0.1600\,t\ \text{per seam site}.
\end{equation}
The circumference dependence is not negligible: at fixed $L_y=20$, $\tau=0.16084$, $0.16004$, $0.15924$, $0.15887$, $0.15867\,t$ for $L_x=20,24,32,40,48$, and a fit $\tau_\infty+a/L_x^2+b/L_x^4$ gives $\tau_\infty=0.15822t$. This is an extrapolation in the circumference at fixed width; repeating it at $L_y=12,16,24,32$ gives $\tau_\infty=0.15825$, $0.15822$, $0.15822$, $0.15822\,t$, so the width dependence of the extrapolated value is below $10^{-5}t$ for $L_y\ge16$ and $0.158t$ is our estimate of the two-dimensional limit, with that uncertainty; $0.1600t$ is the fixed-circumference value. In either case the quantity is the ground-state energy excess of the fixed pairing background, not the equilibrium tension of a self-consistently reconstructed wall. For the covariant $p+ip$ state the analogous quantity depends on what is compared: a $\cD$-eigenblock of the cover has exactly half the cover energy (difference $6\times10^{-14}$), which is an identity and not a physical statement about a quotient; the literal gluing of the same $p+ip$ state ($\mu=1.3t$) costs $\tau_{\rm same}/\tau_{\rm cover}=0.0975/0.0978\,t$ at $L_y=20$, not zero. The tension is a property of the gluing Hamiltonian, not a function of the Chern number alone.

\section{Discussion}\label{sec:discussion}

\paragraph{What is established.} On the Brillouin Klein bottle, pairing turns the $\mathbb Z_2$ glide into a $\mathbb Z_4$ projective symmetry and adds a second, antiunitary obstruction; in the $(\mu,\mathrm{real}\,\Dd)$ section the odd-Chern phases are two ellipses fixed by four Pfaffians, touching the Klein-bottle line only at the nodal point, and the gap closings we identify are three analytic families, two of which (the lines $\mu=\pm2\sqrt2\,t$ and the segment on $\Dd=\pm\Dz$) a coarse momentum grid misses. On the real-space Klein bottle, the uniform chiral state cannot be lifted, the seam is a $\Delta C=2$ wall, and its spectrum, resolved on a translation-invariant unfolding, is that of two co-propagating chiral Majorana branches, degenerate throughout the resolved in-gap window, crossing at the seam Fermi momentum---a chiral Dirac fermion of velocity $2t$---with the level formula \eqref{eq:tower2} reproduced to $10^{-8}$, a finite-size near-zero pair in the sectors where the Fermi momentum is allowed, a tension of $0.1600t$ per site at circumference $24$ ($0.158t$ extrapolated in the circumference), and additional seam-bound levels that the cylinder lacks. These are model-specific, reproducible statements, and they agree with the general picture of Refs.~\cite{Beugeling2014,Quelle2016,Serban2010} where they overlap.

\paragraph{Scope and limitations.} The near-zero pair is not symmetry-protected: it exists only in the sectors selected by circumference and holonomy, it is lifted at first order by specific local seam terms at each tested width and at second order by others, and no symmetry of the tested lattice type beyond $T_{L_x/2}$ (for $L_x\equiv0\bmod4$) and $\cC$ exists. It is therefore not a qubit, and we make no claim about topological quantum computation. The $\cD$-eigenblocks of the cover are not two Pin quotients, and a canonical fermionic quotient for the covariant state remains to be constructed. The seam is a physical domain wall with a localized, gauge-invariant in-gap spectrum, unlike the reflection defect of the gapless staggered fermion, whose location is a gauge choice \cite{SeibergZhang2026}; but we have not proved that the wall ``cannot be moved by any unitary'', and we do not claim it. Whether explicit cross-couplings of two copies can realize the seam is left open. What the non-orientable gluing does, at the level of the free BdG problem, is to close a cylinder with two chiral edges into a surface with one wall, at the cost of a tension and with a small set of seam-bound levels; it does not manufacture new gapless degrees of freedom, and the interesting differences from the cylinder are the ones that depend on the gluing: the holonomy sectors, the seam-bound levels, the mirror-locality of entanglement cuts.

\paragraph{What the pairing diagnostic adds.} The finite odd-frequency response connects the near-zero spectrum, its spatial weight and its response to allowed perturbations. It does not identify the Klein topology by itself: both the cut edges and ordinary walls have a signal, and the relative amplitudes depend on the selected channel. Our comparisons use $\omega/t\ge10^{-3}$; no numerical zero-frequency pole or new size-scaling law is inferred. The static pairing field is not determined self-consistently. Moreover, lifting a finite near-zero pair is not a demonstration that mixing two co-propagating branches gaps the entire chiral interface. Protected storage, coherent transfer and probe readout require distinct encoding, dynamical or measurement models. The present spectrum and correlations establish none of these functions or a universal gate set.

\paragraph{Two antagonistic $\mathbb Z_2$'s.} The non-orientability of the Klein bottle and the particle--hole structure of a Majorana mode play distinct roles. The Klein-bottle $\mathbb Z_2$ is the class of a \emph{commuting spatial} symmetry generated by the projective translation algebra; it supports the $\mathbb Z_2$ invariant of Appendix~\ref{app:Z2} in the normal state and confines $C$ through Theorem~\ref{thm:confine}. The Majorana $\mathbb Z_2$ is the \emph{anticommuting internal} particle--hole structure, exempt from the theorem, which supports $C\in\mathbb Z$ and the Majorana branches of the wall. The two are woven into one algebra by $\cA=(-1)^FR\Gt$ and $[\cC,\Gt]=0$, where they occupy antagonistic positions: the orientation-reversing structure that makes the Brillouin zone a Klein bottle forbids the net-chirality interface that these Majorana branches require, and on the real-space Klein bottle the conflict resolves itself by force, with a wall and a tension. For Majorana-based computation the Klein-bottle Brillouin zone is a selection rule, not a protection mechanism; the real-space Klein bottle is a source of a chirality wall, not of a protected zero mode. Cross-cap defects have been shown to implement fault-tolerant logical gates in stabilizer codes \cite{Kobayashi2024crosscap}; whether a free-fermion chirality wall on a closed non-orientable surface can be turned into a comparable resource is a question for interacting or gauged versions of the model, not for this paper.

\paragraph{Relation to the mod-8 anomaly.} The staggered fermion on the Klein bottle carries a mod-8 parity/time-reversal anomaly derived from the projective algebra of its zero modes \cite{SeibergZhang2026}, and free Majorana fermions on lattice Klein bottles realize the $\mathbb Z_8$ Arf--Brown--Kervaire invariant \cite{ArakiABK2025}. Our model sits on the chiral side of this ledger: the pairing that opens $C=-1$ breaks reflection and time reversal, so the reflection-protected $\mathbb Z_8$ classification does not apply to the bulk; what non-orientability enforces here is the Chern-orientation obstruction, realised locally by the chirality wall. We have not attempted an anomaly-matching computation relating the wall to the $\mathbb Z_8$ invariants, and we do not claim one. The interplay of glide-related copies with interactions remains open. A related gauged model is studied by Junhao (Eigen) Wang in the companion preprint \emph{An exactly solvable obstructed seam on a lattice Klein bottle: bulk-locked Majorana rings and register-dependent toric codes}~\cite{Wang2026paper3}; the present results do not depend on that manuscript.

\paragraph{Measurability.} The BdG Hamiltonian is a single-particle matrix. The tower \eqref{eq:tower2}, its holonomy shift and the seam-bound levels are accessible to classical-wave platforms that emulate lattice matrices with programmable connectivity, of the kind that realized the acoustic Brillouin Klein bottle; a real-space Klein bottle is a wiring pattern. The imaginary-frequency anomalous correlations computed here are not conductance data or a measurement instrument; connecting them to an observable requires a specified probe and response calculation.

\paragraph{Outlook.} Open problems that this paper makes concrete are: a canonical fermionic quotient for glide-covariant pairing on the lattice Klein bottle; an effective seam theory that explains which bilinears lift the periodic-sector pair (the first-order lifting terms found here all carry the seam's Fermi-point structure, and a selection rule for them should be derivable); the self-consistent shape of the wall under a dynamical order parameter; and the spectrum of walls on surfaces with several cross-caps.

\section*{Acknowledgments}
The author designed and directed the research. Anthropic Claude assisted with code, figures and manuscript drafting; OpenAI GPT/Codex assisted with critical checks, the finite odd-frequency calculation and its audit, figures, and revision drafting. The AI-assisted checks do not constitute independent human peer review. Historical reports and responses remain in the original research archive; the public package contains code, results and reproduction records (Appendix~\ref{app:code}). The author takes full responsibility for the content. This work received no external funding.

\appendix

\section{Methods and cross-checks}\label{app:methods}
\emph{Chern numbers.} Fukui--Hatsugai--Suzuki lattice field strengths of the two negative-energy BdG bands on a $72\times144$ momentum grid ($k_x\in[0,\pi)$, $k_y\in[0,2\pi)$) \cite{FukuiHatsugaiSuzuki2005}, evaluated only where the continuous-momentum gap exceeds $0.02t$; on the $81\times81$ parameter grid the unrounded result is an integer to machine precision at all $6422$ such points (maximal deviation $2\times10^{-15}$; an earlier version quoted $0$, an artefact of rounding before the audit), with $2928$ points at $C=0$, $1229$ at each of $C=\pm1$ and $518$ at each of $C=\pm2$. \emph{Gaps.} $\min_\ka|E|$ on a $97\times193$ grid, followed by five rounds of a shrinking $41\times41$ local search started from each of the four lowest distinct grid minima, the four particle--hole-invariant momenta evaluated exactly, and a $4001$-point scan of the line $k_y=\pi$ on which the nodal segment lies (a single-start search can follow the wrong branch where two branches of the minimum cross, as happened at the $\mu=0$ optimum in an earlier version); the three analytic node families are evaluated directly. \emph{Pfaffians.} The analytic Pfaffians of Prop.~\ref{prop:pf} decide the parity at every grid point and are compared with the FHS parity (no mismatch); independently, the numerical Pfaffian of the $4\times4$ antisymmetric $H(\mathbf K)\tau_x$ is evaluated at all $6561$ grid points and agrees with the formula to $10^{-14}$ (\texttt{p1\_node\_scan.py}). \emph{Node scan.} A $161\times161$ scan of the $(\mu,\Dd)$ section (spacing $0.05t$) with a $65\times129$ momentum grid flags every point whose grid gap is below $0.06t$ ($1630$ points); each flagged point is then refined with the multi-start minimizer. All $94$ refined closings (gap below $10^{-3}t$) lie within $7\times10^{-4}t$ of one of the three analytic families (the largest offsets are the grid points $(\pm2,\pm0.975)$ and $(\pm2,\pm1.025)$ next to the tips of the ellipses, with gaps $6\times10^{-4}t$). The $86$ flagged points farther than $0.06t$ from every family all lie in the strip $2.05\le|\mu|\le2.75$, $0.875\le|\Dd|\le1.15$ around the nodal segment (C), where the gap opens slowly, and their refined gaps are $0.027$--$0.058t$: none is a closing. This supports, without proving, that the displayed families are the only gap closings of the section (\texttt{p1\_node\_scan.py}). \emph{Bott index.} $\mathrm{Bott}=\tfrac1{2\pi}\sum_j\arg\lambda_j$ for the eigenvalues $\lambda_j$ of $W_xW_yW_x^\dagger W_y^\dagger$, $W_a=P\,e^{2\pi i a/L_a}P$ restricted to the negative-energy projector $P$ of the real-space BdG torus \cite{LoringHastings2010}; Table~\ref{tab:bott}. \emph{Klein bottle.} Explicit quotient (construction a) and covering-torus deck projection (construction b), spectra agreeing to $10^{-14}$ in the normal state; the cylinder control multiplies every cross-seam amplitude by zero; the antiperiodic holonomy is a flat $\mathbb Z_2$ cochain with $-1$ on the $x$ bonds of one column and compensating signs on the cross-seam bonds, checked plaquette by plaquette (on the cylinder, a sign on the $x$ and diagonal bonds of one column). \emph{Seam dispersion.} The unfolded ribbon of Sec.~\ref{sec:KB}: cut the bonds between rows $L_y/2-1$ and $L_y/2$, reflect rows $y\ge L_y/2$ by $x\to1-x$, verify $T_2$ invariance (residual $0$), and Bloch-reduce with the two-site cell (site momentum $q$, cell phase $e^{2iqm}$; $L_x=8$, $L_y=48$; seam window: three rows on either side); branches are tracked by sign of energy. The Fourier transform of closed-system eigenvectors along $x$ is used only as a spectral-weight diagnostic. \emph{Entanglement.} Majorana covariance matrices \cite{Peschel2003}; the torus reference is subtracted; cuts are mirror-symmetric; the fit uses strips with $4\le a\le L_x-4$. \emph{Symmetry algebra.} Elements represented as (matrix, antiunitary flag, affine $\ka$-map) triples, closed by breadth-first multiplication, with exact matrices (phases $\pm1$) and modulo scalar phases; the action on $H(\ka)$ is tested at a generic $\ka$ for each element. \emph{Real-space symmetry search.} All compositions listed in Sec.~\ref{sec:KB} on $8\times6$; residual threshold $10^{-9}$. The baseline numbers described in this appendix are produced by scripts of the version~1.0.0 reproduction package (Appendix~\ref{app:code}), with sizes, boundary conditions and seeds recorded in its output files.

\begin{table}[!t]
\caption{Bott index of the negative-energy projector on $L\times L$ tori against the FHS Chern number of the Bloch model.}
\label{tab:bott}
\centering\footnotesize
\begin{tabular}{lcccc}
\toprule
$(\mu,\Dd)/t$ & gap$/t$ & $C_{\rm FHS}$ & Bott$_{16}$ & Bott$_{24}$\\
\midrule
$(0,1)$ & $1.5686$ & $-1$ & $-1.0000$ & $-1.0000$\\
$(0,0.3)$ & $0.6000$ & $-1$ & $-1.0000$ & $-1.0000$\\
$(1,-1)$ & $1.0000$ & $+1$ & $+1.0000$ & $+1.0000$\\
$(0.5,1.9)$ & $0.1318$ & $-1$ & $-1.0000$ & $-1.0000$\\
$(2,2)$ & $0.6858$ & $-2$ & $-2.0000$ & $-2.0000$\\
$(3,1.5)$ & $0.1716$ & $0$ & $0.0000$ & $0.0000$\\
$(4,1)$ & $1.1716$ & $0$ & $0.0000$ & $0.0000$\\
$(1.3,0)$ & $0.3854$ & $0$ & $0.0000$ & $0.0000$\\
$(0,i)$ & $0.8284$ & $0$ & $0.0000$ & $0.0000$\\
$(0,e^{i75^\circ})$ & $0.4350$ & $0$ & $0.0000$ & $0.0000$\\
\bottomrule
\end{tabular}
\end{table}

\section{Klein-bottle \texorpdfstring{$\mathbb Z_2$}{Z2} invariant and nonlocal edge sewing}\label{app:Z2}
\begin{figure}[!t]
\centering\includegraphics[width=\columnwidth]{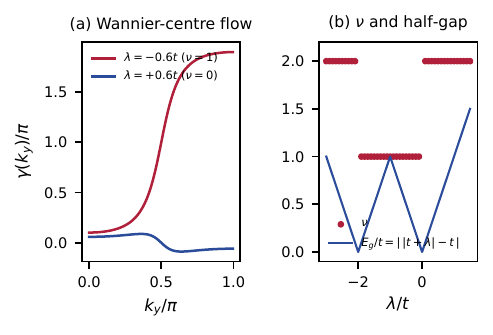}
\caption{(a) Wannier-centre flow $\gamma(k_y)$ (continuous branch) over half the Brillouin zone for $\lambda=-0.6t$ ($\nu=1$) and $+0.6t$ ($\nu=0$). (b) $\nu$ and the half-gap $\big||t+\lambda|-t\big|$ versus $\lambda$.}
\label{fig:zak}
\end{figure}

\begin{figure}[!t]
\centering\includegraphics[width=\columnwidth]{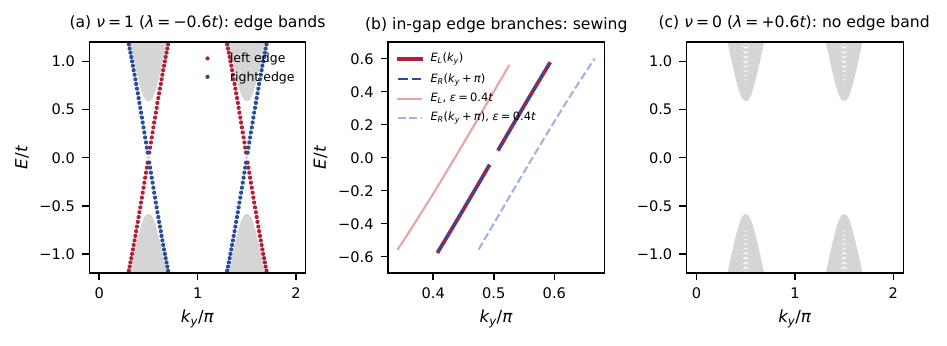}
\caption{Nonlocal edge sewing on a 60-cell strip open in $x$: (a) $\nu=1$ ($\lambda=-0.6t$) edge bands, left edge red, right edge blue; (b) the in-gap branches $E_{\rm L}(k_y)$ and $E_{\rm R}(k_y+\pi)$ coincide to $4\times10^{-15}$; with a $G_2$-breaking sublattice potential $\epsilon=0.4t$ (pale curves) they separate by $0.8t$; (c) $\nu=0$ ($\lambda=+0.6t$): no edge band.}
\label{fig:edge}
\end{figure}

The half-filled normal state is a Dirac semimetal; the $\mathbb Z_2$ classification of Ref.~\cite{ChenYangZhao2022} concerns insulators. A columnar dimerization of the $x$ bonds, $t+\lambda$ (intra-cell) and $t$ (inter-cell), adds $-\lambda\sigma_x$ to Eq.~\eqref{eq:H0}, preserves $G_2$ and time reversal, and opens a gap $E_g=\big||t+\lambda|-t\big|$ (half the band gap), which closes at $\lambda=0$ and $\lambda=-2t$: $\lambda=-0.6t$ gives $0.6t$, $\lambda=-1.5t$ gives $0.5t$. With $\gamma(k_y)$ the Zak phase of the occupied band along $k_x$ (continuous branch, fixed unit-cell convention), the glide implies $\gamma(k_y)+\gamma(k_y+\pi)=0\bmod2\pi$ (verified to $2\times10^{-15}$; broken by a $G_2$-breaking sublattice potential, residual $1.9$ at $\epsilon=0.4t$) and $\nu=\tfrac1{2\pi}[\gamma(0)+\gamma(\pi)]\bmod2$; on the lines $k_y=\pi/2,3\pi/2$, where the $\sigma_z$ term of Eq.~\eqref{eq:H0} vanishes and each line is an SSH chain with sublattice chiral symmetry, the Zak phase is quantized, $\gamma(\pi/2)=\pi$ for $\lambda=-0.6t$ and $0$ for $\lambda=+0.6t$; on $k_y=0,\pi$ it is not ($\pm0.3185$ for $\lambda=-0.6t$), and $\nu$ is defined through the continuous branch, not through a single-line quantization. In a fixed unit-cell convention $\nu=1$ for $\lambda\in(-2t,0)$ and $\nu=0$ otherwise (Fig.~\ref{fig:zak}); $\nu$ flips under a one-site shift of the cell, so only the difference of the two phases, the transition points, and the sewing relation below are convention-independent. The Majorana platform's normal state $\lambda=0$ is exactly the critical point between the two Klein-bottle classes. On a strip open in $x$, the $\nu=1$ phase has one edge band per edge crossing the gap at $k_y=\pi/2,3\pi/2$, and the two macroscopically separated edges are sewn pointwise, $E_{\rm L}(k_y)=E_{\rm R}(k_y+\pi)$, to $4\times10^{-15}$ (Fig.~\ref{fig:edge}); a $G_2$-breaking sublattice potential $\epsilon=0.4t$ destroys the sewing (mismatch $0.8t$). This reproduces the nonlocal twist of Ref.~\cite{ChenYangZhao2022} in the present model.


\section{Glide bilayer: translation, not glide, protects the edge}\label{app:bilayer}
\begin{figure*}[!t]
\centering\includegraphics[width=0.8\textwidth]{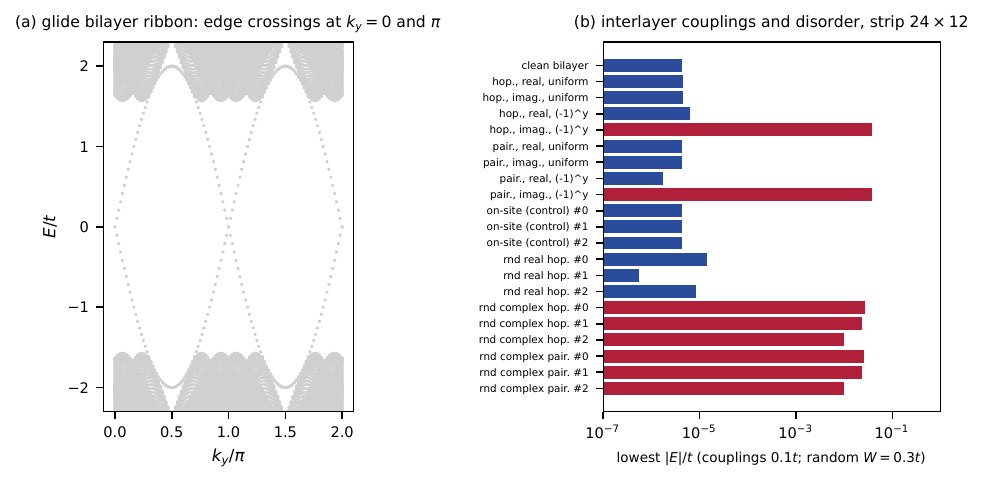}
\caption{Glide bilayer. (a) Ribbon spectrum ($40$ cells, Bloch $k_y$): edge crossings at $k_y=0$ and $\pi$. (b) Interlayer-coupling ledger on the $24\times12$ strip (red: lowest $|E|$ above $10^{-3}t$): the uniform and $(-1)^y$ couplings of strength $0.1t$, the intralayer on-site control ($W=0.3t$, three realizations) and random interlayer couplings on the edge columns ($W=0.3t$, three realizations each). The imaginary $(-1)^y$ terms and the random complex interlayer terms lift the edge states; real couplings, uniform, staggered or random, do not at this resolution.}
\label{fig:bilayer}
\end{figure*}

Stacking a $C=-1$ layer with its glide image (same $p+ip$, $\Dd$ on the other diagonal, $C=+1$) gives a glide-symmetric bilayer, $\mathcal W=(\text{layer swap})\otimes\Gt$, $\mathcal W^2=(-1)^F$, with $C=0$. On a ribbon each edge carries two counter-propagating Majorana branches whose crossings are displaced by exactly $\pi$ (layer 1 at $k_y=0$, layer 2 at $k_y=\pi$; Fig.~\ref{fig:bilayer}a). On a $24\times12$ strip (open in $x$, periodic in $y$) interlayer couplings of strength $0.1t$ that are uniform along the edge---real or imaginary hopping, real or imaginary pairing---and real couplings with a $(-1)^y$ modulation leave the edge gapless ($\le6\times10^{-6}t$, the clean value being $4\times10^{-6}t$); an \emph{imaginary} $(-1)^y$ interlayer hopping or pairing gaps it at first order ($0.0374t$ and $0.0375t$; Fig.~\ref{fig:bilayer}b). Random on-site disorder within the layers ($0.3t$ on the edge columns, three realizations) also leaves it gapless, but this is a control without content: an intralayer potential does not couple the layers at all. The informative test is random \emph{interlayer} coupling on the edge columns, which breaks translation and contains every momentum and phase. A random \emph{real} interlayer hopping of strength $W=0.3t$ leaves the near-zero edge states at $10^{-6}$--$10^{-5}t$; on a $48$-column strip, where the clean values are $10^{-12}t$, the lift scales as $W^{3}$ (from $10^{-8}t$ at $W=0.05t$ to $5\times10^{-6}t$ at $W=0.4t$, log--log slope $2.99$), and the first-order matrix of the perturbation in the four-dimensional near-zero subspace vanishes ($<10^{-13}$). A random \emph{complex} interlayer hopping or pairing lifts them at first order (slope $0.98$): to $0.003$--$0.03t$ at $W=0.1$--$0.3t$ on $24\times12$, with realization-dependent values at $L_y=24,48$ because the first-order term is the momentum-$\pi$ imaginary component of the particular pattern. The protection is therefore that of the pair of weak Kitaev $\mathbb Z_2$ indices at $k_y=0,\pi$ (translation and particle--hole symmetry) against translation-preserving couplings only: the glide forces the two crossings apart by $\pi$ and the two weak indices to be equal, the term that gaps the edge is the momentum-$\pi$ term with an imaginary phase, and generic interlayer disorder, which contains it, lifts the edge states at first order. The insensitivity to real couplings of any spatial pattern is a selection rule of this model, not a symmetry: a search over $2240$ lattice operations (seven point-group and translation elements, ten gauge patterns, four layer and four Nambu matrices, unitary or antiunitary) finds $28$ symmetries and spectral antisymmetries of the clean bilayer, of which only the identity and particle--hole conjugation survive a random real interlayer hopping. Weak indices multiply under stacking, so two bilayers already cancel the indices analysed here; we make no statement about other crystalline classifications. A vortex core with $N$ Majoranas keeps one at zero energy for odd $N$, while for even $N$ generic symmetry-allowed mixing lifts all of them.

\input{odd_methods}

\section{Code, data and reproducibility}\label{app:code}
The accompanying release is titled \emph{Paper 1 reproducibility package: Chiral superconductivity on a closed Klein bottle}. The baseline version 1.0.0, preceding the odd-frequency extension, is archived on Zenodo at \href{https://doi.org/10.5281/zenodo.23476917}{doi:10.5281/zenodo.23476917}. The article text, original figures, package documentation and original result data use CC BY 4.0; original reproduction code uses the MIT License. Third-party dependencies retain their own licences.

The combined reproduction archive, version 1.1.0, separates the preserved baseline, the implementation-validation history and the registered odd-frequency extension into \texttt{baseline/}, \texttt{validation/} and \texttt{extension/}. The original relative layout is retained beneath \texttt{baseline/evidence/project/}. The baseline scripts include the phase, symmetry, Bott-index, covering-space, seam, unfolded-ribbon, wall, entanglement, bilayer, normal-state and node-scan calculations, with their model builders and saved results. No separate project checkout is required.

The combined package entry point is \texttt{run.py}, using an explicitly supplied interpreter. Its \texttt{read} mode extracts saved extension numbers; \texttt{validate} runs the archived implementation checks; \texttt{quick} runs a documented subset; \texttt{full} replays all registered extension configurations and their analysis; and \texttt{figures} regenerates plots from saved or replayed results. Baseline read and quick modes are selected separately; the original complete baseline chain remains available through \texttt{baseline/run.py}. Its historical estimate is about ninety minutes on one core, and it was not repeated for this revision. The package documents precisely which original and new calculations were replayed, rather than treating saved-data reads as recomputation.

The extension record includes registration, configuration identifiers, complex component arrays, working error diagnostics, complete-subspace spectral summaries and plotting inputs. Local acceptance uses an archive extracted into a fresh directory and a newly installed dependency environment, with CPython~3.13.5, NumPy~2.1.3 and Matplotlib~3.10.0. The interpreter, binary-library configuration, dependency hashes, tolerances, elapsed times, memory use and cache policy accompany the replay records. These are local checks on the same host, not validation on every platform. A separate implementation checks representative configurations and finite-frequency formulas; its limited coverage and verifier-only dependencies are reported explicitly. AI-assisted checks do not replace independent human review.

The manuscript and supplement source bundles include their figures and the matching compiled PDFs; the main bundle also includes \texttt{references.bib} and \texttt{main.bbl}. Historical absolute paths in frozen logs identify the original runs; the supported combined entry points use package-relative inputs and new temporary output directories. The original development archive remains unchanged.

The quoted working errors and tabulated resolution flags refer to the frozen reference output. A fresh offline installation of the exact reference package builds reproduces all 23 configurations and all 308 comparison-run lists, including the printed values and flags. The actual runtime libraries and numerical NumPy extensions are checked by hash before reference computations; supported architecture, kernel and thread settings are declared explicitly. The \texttt{wp6r/} records separate this reproduction from the failed original Accelerate cross-build test and the post hoc shared-margin sensitivity check of Appendix~\ref{app:odd-builds}. The latter does not turn working estimates into certificates or establish invariance on untested systems. The complete difference ledger and the original failure remain available.

The version~1.0.0 DOI above does not archive the odd-frequency extension. The combined reproduction archive, version~1.1.0, is identified by \href{https://doi.org/10.5281/zenodo.23967897}{doi:10.5281/zenodo.23967897} and includes the extension and local acceptance records. The supplement accompanies this revision.

\bibliographystyle{unsrtnat}
\bibliography{references}
\end{document}

\maketitle
\section*{Scope and how to read the data}
This supplement reports the finite imaginary-frequency extension to the model of arXiv:2610.40251v1. It accompanies the revised manuscript and reproduction archive version~1.1.0. The main manuscript gives the resolvent definitions, Pauli relation, complete spectral-projector formulas, geometry windows and numerical error prescription. Here we preserve the four-channel plots, both ordinary walls, both perturbation signs, and both registered spectral cutoffs.

The registration is \texttt{P1-ODD-FREQ-20261102-v1}. All 23 main configurations were computed; no conditional size or trivial-phase extension was activated. The principal discussion is restricted to $0.001\le\omega/t\le10$. Lower-frequency values remain visible with their resolution flags; additional direct checks required for interpreting that range were not performed. No curve is fitted to a power law. Unresolved is not zero, and lack of a resolved difference is not equality. The working errors are numerical diagnostics rather than statistical errors or rigorous interval enclosures.

The model uses $t=\Delta_0=\operatorname{Re}\Delta_{\rm d}=1$, $\mu=\lambda=\operatorname{Im}\Delta_{\rm d}=0$, before the stated wall texture or perturbation. All systems have one spinless layer. The $x$ twist changes holonomy, not the checked local plaquette flux. The ordinary-wall torus has sign-reversed diagonal pairing for source rows $y\ge24$.

\begin{center}\small
\begin{tabular}{lll}\toprule
IDs & Geometry and size & Change from clean periodic model\\\midrule
M01, M02 & Klein $24\times24$ & P, $\mathrm{AP}_x$\\
M03, M04 & cut cylinder $24\times24$ & P, $\mathrm{AP}_x$\\
M05 & uniform torus $24\times24$ & none\\
P00 & Klein $24\times32$ & none\\
PS01--PS08 & Klein $24\times32$ & $a(-1)^x$ vertical pairing, row 0\\
PB01--PB08 & Klein $24\times32$ & same amplitudes, row 16\\
W01 & two-wall torus $24\times48$ & diagonal-pairing sign texture\\\bottomrule
\end{tabular}\end{center}
For both PS and PB, indices 01--08 correspond respectively to
$a/t=-0.2,-0.1,-0.05,-0.025$, followed by $+0.025,+0.05,+0.1,+0.2$.
The pair sets are normalized separately; no sum over two walls is compared to a single seam. Cross-interface bonds and sublattice groups are retained in the underlying calculation record, with absent cylinder interfaces marked not applicable.

\paragraph{Reproduction record.} The baseline archive, \href{https://doi.org/10.5281/zenodo.23476917}{doi:10.5281/zenodo.23476917}, does not contain this extension. The combined reproduction archive, version~1.1.0, is identified by \href{https://doi.org/10.5281/zenodo.23967897}{doi:10.5281/zenodo.23967897} and supplies code, arrays, figure/table inputs, an installable reference-build lock and separate replay records. All values, working errors and figure markers below retain the frozen reference output. Cross-build differences and the post hoc shared-margin check are reported at the end of this supplement; they do not certify numerical bounds. AI-assisted checks are not human peer review.

\clearpage
\section*{Four-channel geometry comparison}
\begin{figure}[!ht]
\centering\includegraphics[width=\textwidth,height=0.70\textheight,keepaspectratio]{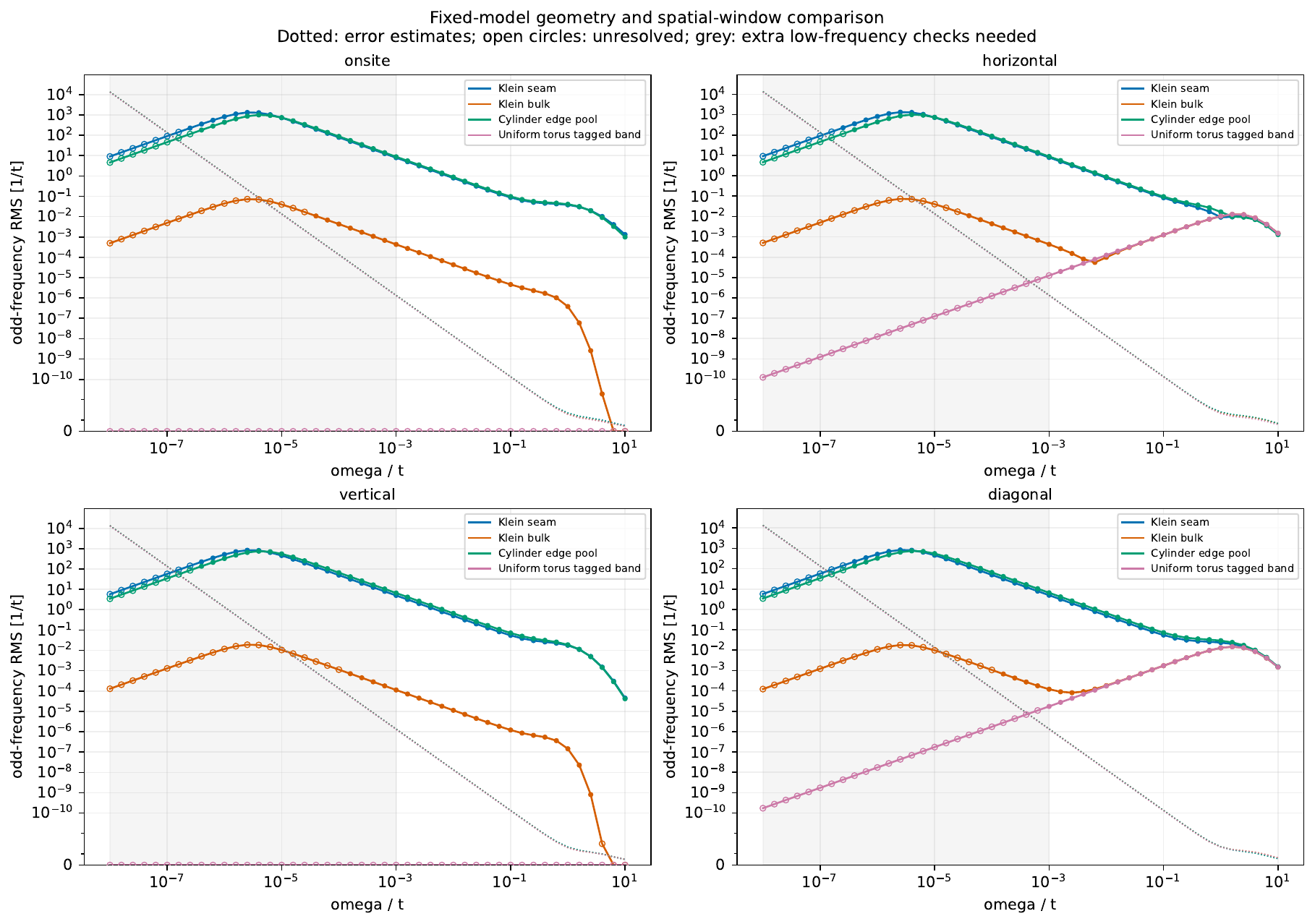}
\caption{Periodic $24\times24$ Klein, cylinder and uniform-torus windows. The torus labelled band is not a wall. Dotted lines show the working numerical errors, not statistical confidence intervals. Hollow markers denote unresolved values. The shaded range $\omega/t<10^{-3}$ is retained for completeness, without the additional direct checks required for a physical low-frequency interpretation.}
\end{figure}

\clearpage
\section*{Holonomy controls}
\begin{figure}[!ht]
\centering\includegraphics[width=\textwidth,height=0.70\textheight,keepaspectratio]{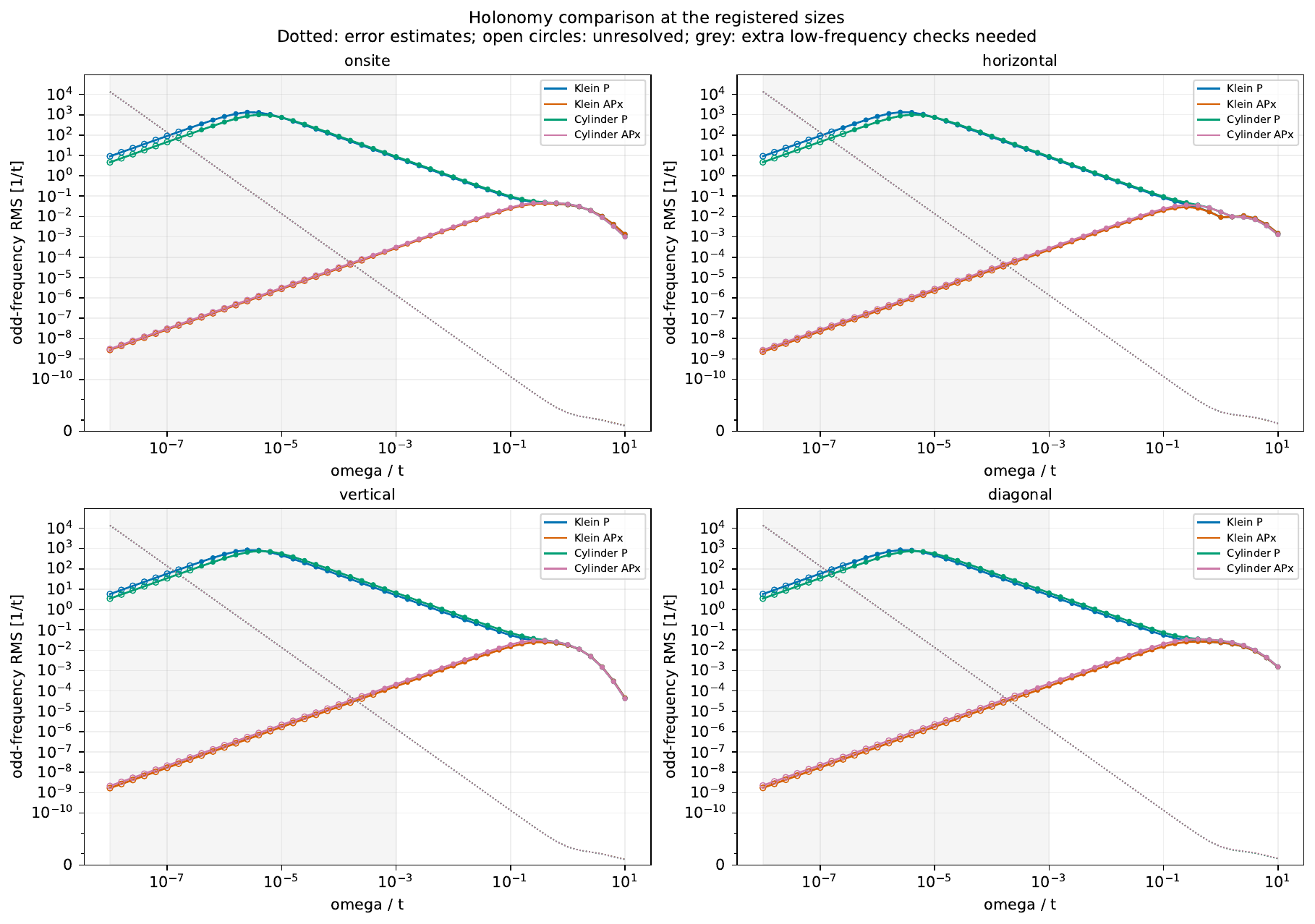}
\caption{Periodic and antiperiodic $x$ holonomy for the $24\times24$ Klein bottle and cut cylinder, in all four channels. These twists preserve the checked local plaquette flux. A common absolute-frequency grid is used. The shaded lower-frequency range is not used for physical conclusions.}
\end{figure}

\clearpage
\section*{Ordinary-wall controls}
\begin{figure}[!ht]
\centering\includegraphics[width=\textwidth,height=0.70\textheight,keepaspectratio]{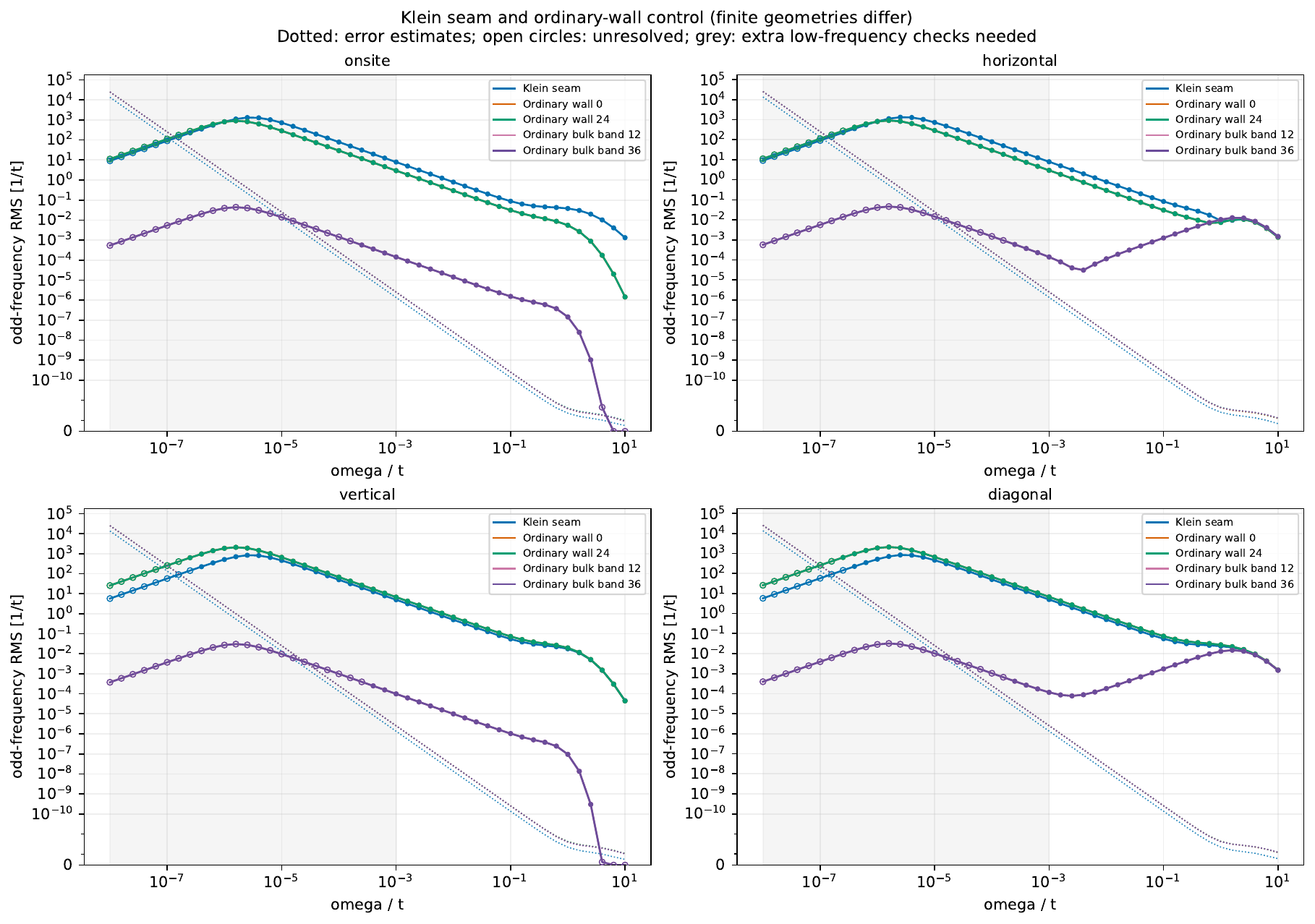}
\caption{The $24\times48$ torus changes the sign of the source-row diagonal pairing for $y\ge24$. Both six-row wall bands and both bulk bands are separately normalized and displayed. Klein values are included as a model comparison, not an isolated test of non-orientability. The local templates and total sizes differ. This finite $y$-wall configuration is distinct from the earlier Bloch-wall reference in the main article.}
\end{figure}

\clearpage
\section*{Four-channel spatial profiles}
\begin{figure}[!ht]
\centering\includegraphics[width=\textwidth,height=0.70\textheight,keepaspectratio]{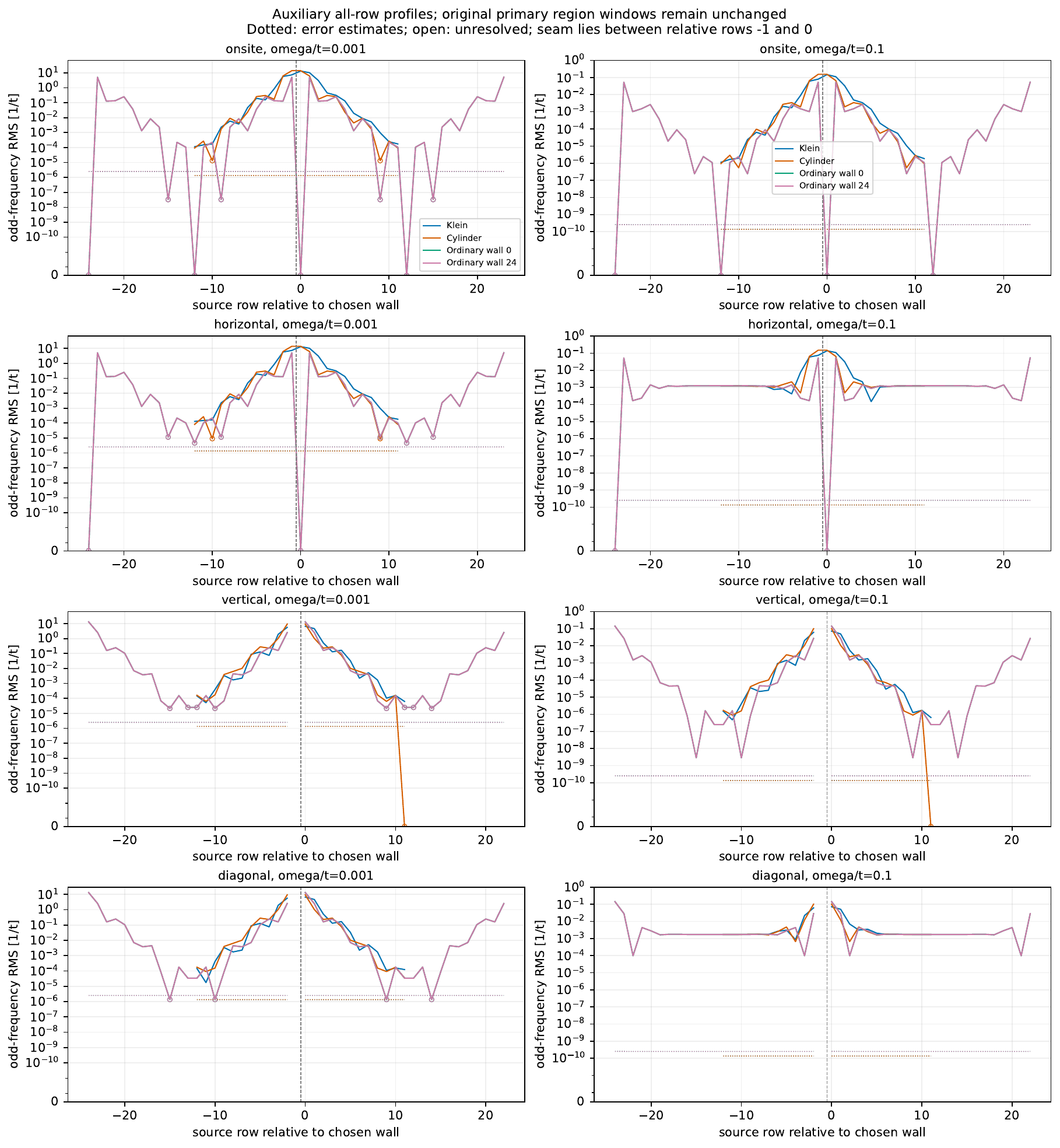}
\caption{Auxiliary all-row profiles at $\omega/t=0.001$ and $0.1$ use the same component definitions, without changing the preregistered band statistics. The four rows show onsite, horizontal, vertical and diagonal RMS, respectively; these are correlation profiles, not projector-localization plots.}
\end{figure}

\clearpage
\section*{All registered signed perturbations}
\begin{figure}[!ht]
\centering\includegraphics[width=\textwidth,height=0.70\textheight,keepaspectratio]{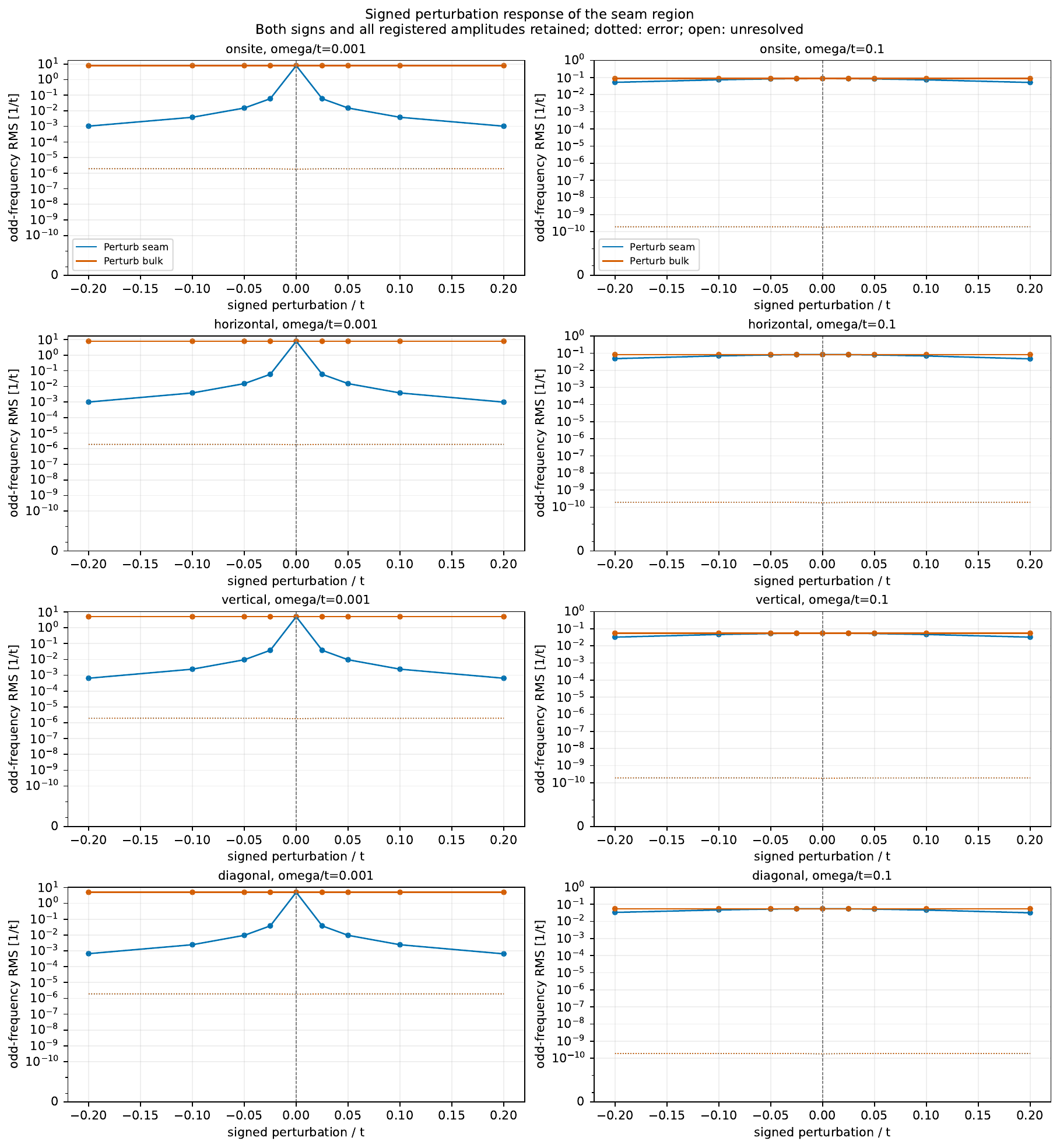}
\caption{Real staggered vertical pairing on source row $0$ (seam) or $16$ (bulk) of the $24\times32$ Klein system. All eight nonzero signed amplitudes and the common clean point are retained. Four-channel seam responses are shown at the fixed frequencies indicated. Apparent overlap does not mean exact invariance; lines do not constitute a power-law fit.}
\end{figure}

\clearpage
\section*{Resolution and numerical diagnostics}
\begin{figure}[!ht]
\centering\includegraphics[width=\textwidth,height=0.70\textheight,keepaspectratio]{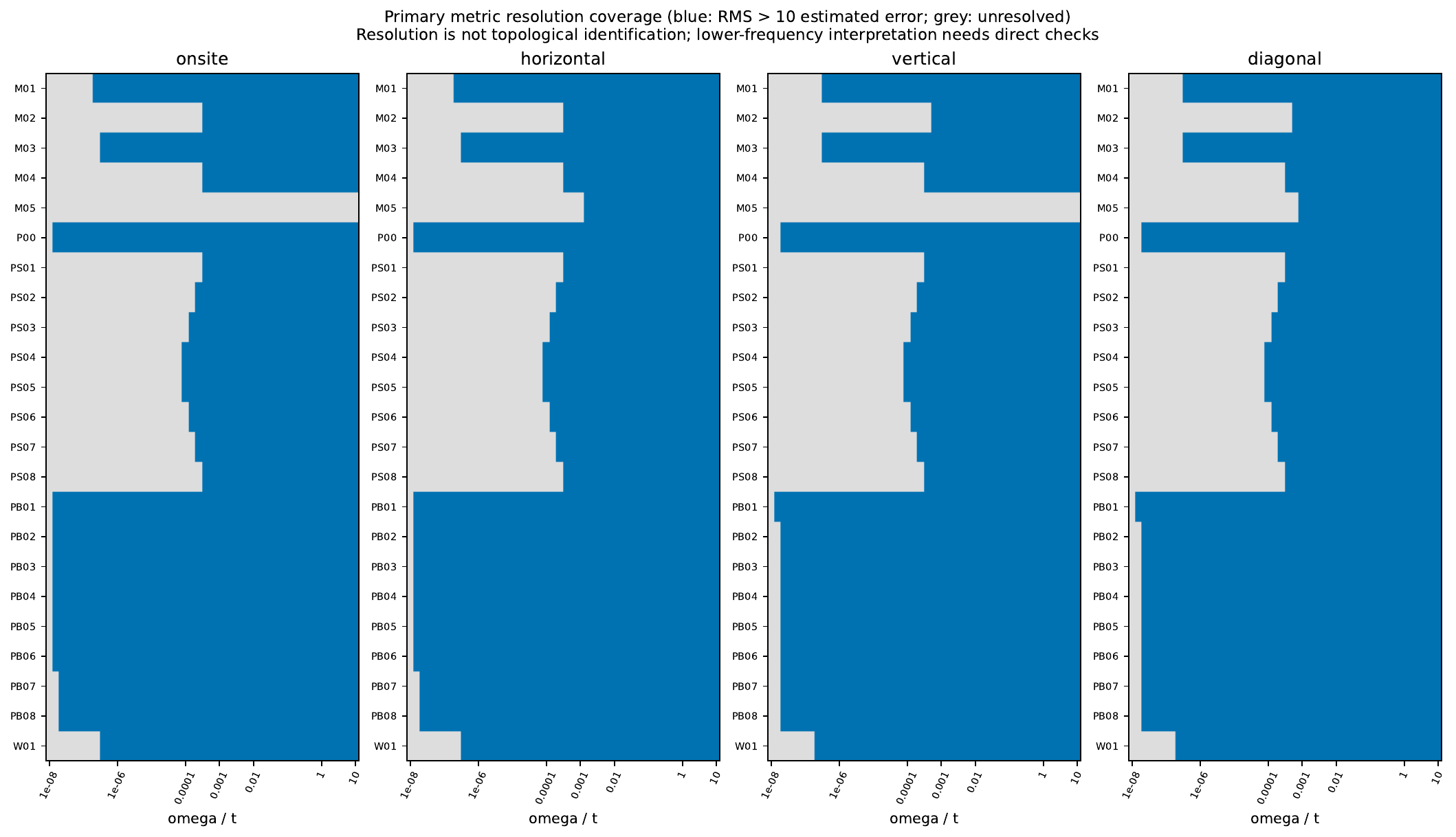}
\caption{Coverage panels show the 46 standard positive frequencies for all 23 configurations in each of the four channels. They use the seam or pooled-edge/labelled window, except W01 uses wall 0, and distinguish resolved from unresolved RMS under the working-error prescription. The 202 accepted gap-scaled additions are retained in the underlying data but are not plotted here. This is a coverage map, not a residual or cutoff-guard plot. Lower-frequency entries are not a numerical determination of a zero-frequency pole or asymptote.}
\end{figure}
\clearpage\section*{Complete-subspace split at $|E|<0.1t$}
All entries below are on-site band RMS values at $\omega/t=10^{-3}$ (dimensionless $tM$), with $t^2I$ the interference term. The identity is $M^2=L^2+R^2+I$, not $M=L+R$. The tables retain the sign of $I$. Rank counts BdG eigenstates, including particle--hole partners. A superscript U denotes an unresolved RMS in the frozen reference run; zero from an empty retained subspace is a definition of that decomposition, not a proof of vanishing full correlations. Exact values, estimated errors and flags are supplied in \texttt{table\_data.json}. W01 walls are shown separately; M03/M04 pool the two edges and M05 uses a labelled torus band.
\begin{center}\scriptsize\setlength{\tabcolsep}{4pt}
\begin{tabular}{lrrrrr}\toprule
Case/window & rank & $tM_{\rm full}$ & $tM_{\rm low}$ & $tM_{\rm rest}$ & $t^2I$\\\midrule
M01 & 2 & $7.8938\times10^{0}$ & $7.8937\times10^{0}$ & $9.0025\times10^{-5}$ & $1.4034\times10^{-3}$\\
M02 & 0 & $2.7939\times10^{-4}$ & $(0)^{\mathrm{U}}$ & $2.7939\times10^{-4}$ & $0$\\
M03 & 2 & $8.7756\times10^{0}$ & $8.7755\times10^{0}$ & $9.8571\times10^{-5}$ & $1.7141\times10^{-3}$\\
M04 & 0 & $3.0934\times10^{-4}$ & $(0)^{\mathrm{U}}$ & $3.0934\times10^{-4}$ & $0$\\
M05 & 0 & $(5.5238\times10^{-20})^{\mathrm{U}}$ & $(0)^{\mathrm{U}}$ & $(5.5238\times10^{-20})^{\mathrm{U}}$ & $0$\\
P00 & 2 & $7.8939\times10^{0}$ & $7.8938\times10^{0}$ & $9.0025\times10^{-5}$ & $1.4034\times10^{-3}$\\
PS01 & 2 & $1.0402\times10^{-3}$ & $9.3801\times10^{-4}$ & $1.0346\times10^{-4}$ & $1.9141\times10^{-7}$\\
PS02 & 2 & $3.8417\times10^{-3}$ & $3.7485\times10^{-3}$ & $9.4449\times10^{-5}$ & $6.9863\times10^{-7}$\\
PS03 & 2 & $1.5061\times10^{-2}$ & $1.4970\times10^{-2}$ & $9.1735\times10^{-5}$ & $2.7109\times10^{-6}$\\
PS04 & 2 & $5.9631\times10^{-2}$ & $5.9541\times10^{-2}$ & $9.0761\times10^{-5}$ & $1.0670\times10^{-5}$\\
PS05 & 2 & $5.9629\times10^{-2}$ & $5.9540\times10^{-2}$ & $8.9520\times10^{-5}$ & $1.0529\times10^{-5}$\\
PS06 & 2 & $1.5059\times10^{-2}$ & $1.4970\times10^{-2}$ & $8.9238\times10^{-5}$ & $2.6397\times10^{-6}$\\
PS07 & 2 & $3.8368\times10^{-3}$ & $3.7485\times10^{-3}$ & $8.9327\times10^{-5}$ & $6.6202\times10^{-7}$\\
PS08 & 2 & $1.0293\times10^{-3}$ & $9.3801\times10^{-4}$ & $9.2126\times10^{-5}$ & $1.7109\times10^{-7}$\\
PB01 & 2 & $7.8939\times10^{0}$ & $7.8938\times10^{0}$ & $9.0025\times10^{-5}$ & $1.4034\times10^{-3}$\\
PB02 & 2 & $7.8939\times10^{0}$ & $7.8938\times10^{0}$ & $9.0025\times10^{-5}$ & $1.4034\times10^{-3}$\\
PB03 & 2 & $7.8939\times10^{0}$ & $7.8938\times10^{0}$ & $9.0025\times10^{-5}$ & $1.4034\times10^{-3}$\\
PB04 & 2 & $7.8939\times10^{0}$ & $7.8938\times10^{0}$ & $9.0025\times10^{-5}$ & $1.4034\times10^{-3}$\\
PB05 & 2 & $7.8939\times10^{0}$ & $7.8938\times10^{0}$ & $9.0025\times10^{-5}$ & $1.4034\times10^{-3}$\\
PB06 & 2 & $7.8939\times10^{0}$ & $7.8938\times10^{0}$ & $9.0025\times10^{-5}$ & $1.4034\times10^{-3}$\\
PB07 & 2 & $7.8939\times10^{0}$ & $7.8938\times10^{0}$ & $9.0025\times10^{-5}$ & $1.4034\times10^{-3}$\\
PB08 & 2 & $7.8939\times10^{0}$ & $7.8938\times10^{0}$ & $9.0025\times10^{-5}$ & $1.4034\times10^{-3}$\\
W01 / wall0 & 4 & $2.9486\times10^{0}$ & $2.9486\times10^{0}$ & $(1.8138\times10^{-5})^{\mathrm{U}}$ & $1.0637\times10^{-4}$\\
W01 / wall24 & 4 & $2.9486\times10^{0}$ & $2.9486\times10^{0}$ & $(1.8138\times10^{-5})^{\mathrm{U}}$ & $1.0637\times10^{-4}$\\
\bottomrule\end{tabular}\end{center}
\clearpage\section*{Complete-subspace split at $|E|<1.1t$}
All entries below are on-site band RMS values at $\omega/t=10^{-3}$ (dimensionless $tM$), with $t^2I$ the interference term. The identity is $M^2=L^2+R^2+I$, not $M=L+R$. The tables retain the sign of $I$. Rank counts BdG eigenstates, including particle--hole partners. A superscript U denotes an unresolved RMS in the frozen reference run; zero from an empty retained subspace is a definition of that decomposition, not a proof of vanishing full correlations. Exact values, estimated errors and flags are supplied in \texttt{table\_data.json}. W01 walls are shown separately; M03/M04 pool the two edges and M05 uses a labelled torus band.
A dagger marks a reference U flag that becomes resolved under the original Accelerate working margin. It remains unresolved for both sets of values under the shared-margin check; the number and reference U flag are not replaced.
\begin{center}\scriptsize\setlength{\tabcolsep}{4pt}
\begin{tabular}{lrrrrr}\toprule
Case/window & rank & $tM_{\rm full}$ & $tM_{\rm low}$ & $tM_{\rm rest}$ & $t^2I$\\\midrule
M01 & 10 & $7.8938\times10^{0}$ & $7.8938\times10^{0}$ & $1.9009\times10^{-5}$ & $2.2345\times10^{-4}$\\
M02 & 8 & $2.7939\times10^{-4}$ & $2.5866\times10^{-4}$ & $2.4195\times10^{-5}$ & $1.0572\times10^{-8}$\\
M03 & 10 & $8.7756\times10^{0}$ & $8.7755\times10^{0}$ & $1.8015\times10^{-5}$ & $2.5458\times10^{-4}$\\
M04 & 8 & $3.0934\times10^{-4}$ & $2.8761\times10^{-4}$ & $2.4263\times10^{-5}$ & $1.2384\times10^{-8}$\\
M05 & 0 & $(5.5238\times10^{-20})^{\mathrm{U}}$ & $(0)^{\mathrm{U}}$ & $(5.5238\times10^{-20})^{\mathrm{U}}$ & $0$\\
P00 & 10 & $7.8939\times10^{0}$ & $7.8938\times10^{0}$ & $1.9009\times10^{-5}$ & $2.2345\times10^{-4}$\\
PS01 & 10 & $1.0402\times10^{-3}$ & $1.0252\times10^{-3}$ & $1.9706\times10^{-5}$ & $3.0550\times10^{-8}$\\
PS02 & 10 & $3.8417\times10^{-3}$ & $3.8273\times10^{-3}$ & $1.9204\times10^{-5}$ & $1.0989\times10^{-7}$\\
PS03 & 10 & $1.5061\times10^{-2}$ & $1.5047\times10^{-2}$ & $(1.9069\times10^{-5})^{\mathrm{U},\dagger}$ & $4.2777\times10^{-7}$\\
PS04 & 10 & $5.9631\times10^{-2}$ & $5.9617\times10^{-2}$ & $(1.9030\times10^{-5})^{\mathrm{U},\dagger}$ & $1.6900\times10^{-6}$\\
PS05 & 10 & $5.9629\times10^{-2}$ & $5.9615\times10^{-2}$ & $(1.9007\times10^{-5})^{\mathrm{U},\dagger}$ & $1.6873\times10^{-6}$\\
PS06 & 10 & $1.5059\times10^{-2}$ & $1.5044\times10^{-2}$ & $(1.9023\times10^{-5})^{\mathrm{U},\dagger}$ & $4.2631\times10^{-7}$\\
PS07 & 10 & $3.8368\times10^{-3}$ & $3.8225\times10^{-3}$ & $(1.9110\times10^{-5})^{\mathrm{U},\dagger}$ & $1.0900\times10^{-7}$\\
PS08 & 8 & $1.0293\times10^{-3}$ & $1.0076\times10^{-3}$ & $2.5694\times10^{-5}$ & $4.3522\times10^{-8}$\\
PB01 & 10 & $7.8939\times10^{0}$ & $7.8938\times10^{0}$ & $(1.9009\times10^{-5})^{\mathrm{U},\dagger}$ & $2.2345\times10^{-4}$\\
PB02 & 10 & $7.8939\times10^{0}$ & $7.8938\times10^{0}$ & $(1.9009\times10^{-5})^{\mathrm{U},\dagger}$ & $2.2345\times10^{-4}$\\
PB03 & 10 & $7.8939\times10^{0}$ & $7.8938\times10^{0}$ & $(1.9009\times10^{-5})^{\mathrm{U},\dagger}$ & $2.2345\times10^{-4}$\\
PB04 & 10 & $7.8939\times10^{0}$ & $7.8938\times10^{0}$ & $(1.9009\times10^{-5})^{\mathrm{U},\dagger}$ & $2.2345\times10^{-4}$\\
PB05 & 10 & $7.8939\times10^{0}$ & $7.8938\times10^{0}$ & $1.9009\times10^{-5}$ & $2.2345\times10^{-4}$\\
PB06 & 10 & $7.8939\times10^{0}$ & $7.8938\times10^{0}$ & $(1.9009\times10^{-5})^{\mathrm{U},\dagger}$ & $2.2345\times10^{-4}$\\
PB07 & 10 & $7.8939\times10^{0}$ & $7.8938\times10^{0}$ & $1.9009\times10^{-5}$ & $2.2345\times10^{-4}$\\
PB08 & 10 & $7.8939\times10^{0}$ & $7.8938\times10^{0}$ & $(1.9009\times10^{-5})^{\mathrm{U},\dagger}$ & $2.2345\times10^{-4}$\\
W01 / wall0 & 20 & $2.9486\times10^{0}$ & $2.9486\times10^{0}$ & $(3.6704\times10^{-6})^{\mathrm{U}}$ & $-2.1636\times10^{-5}$\\
W01 / wall24 & 20 & $2.9486\times10^{0}$ & $2.9486\times10^{0}$ & $(3.6704\times10^{-6})^{\mathrm{U}}$ & $-2.1636\times10^{-5}$\\
\bottomrule\end{tabular}\end{center}

\clearpage\section*{Numerical-build sensitivity and reference provenance}
The reference outputs use the declared OpenBLAS runtime. A same-version
Accelerate replay gives closely agreeing physical values but different
working residual estimates. The original test for identical classifications across builds
fails: 21 of 308 qualifying-run lists in $\omega/t\ge10^{-3}$ change,
and the eleven daggered rest entries in the $1.1t$ table change their U
flags. The complete discrepancy ledger accompanies the package. No
alternate-only qualifying comparison is adopted as a new physical result.

The post hoc check recorded before its execution uses
\[
 b_R^{\rm common}=\max(b_R^{\rm ref},b_R^{\rm alt})
\]
at each matched standard-grid scalar. The original strict factor-ten
criterion and three-adjacent-standard-point qualification are applied
separately to the two unaveraged sets of values. Their classifications
then agree across all 308 comparisons, 47,840 primary scalar points and
144 printed flags. Here the maximum coincides with the reference margin.
This empirical check does not establish a rigorous bound, stability on
unseen builds or a low-frequency limit. The original failed test and its
unmodified margins remain archived.

All dotted errors, open/filled markers and coverage panels in Figs.~S1--S6
remain reference-run quantities. The auxiliary spatial records contain
six original flag differences: the cut-cylinder on-site rows $y=9,14$
at $\omega/t=0.001$, and the ordinary-wall vertical rows $y=11,12,35,36$
at that frequency. All are unresolved in the reference output and resolved
with the original alternate margin; both shared-margin classifications
remain unresolved. The primary figure ledger additionally identifies the M05 horizontal
point at $\omega/t=0.001$ in Fig.~S1 and the coverage panel. Lower-frequency
marker changes in the ordinary-wall and coverage figures are retained as
diagnostics only. The complete figure-marker ledger identifies every
selected record; the physical curves and reference flags are not replaced.

The four repeated M05 full/rest entries near $10^{-20}/t$ remain
unresolved. Their printed digits identify the frozen data; different
last digits in the alternate replay have no claimed physical significance.
Cutoff ranks, the printed main physical values and the principal holonomy
and seam/bulk-placement statements retain their stated scope. Exact-build
reproduction and the shared-margin audit are reported separately; neither
is independent human peer review.

%% file: odd_results.tex
\subsection{Finite odd-frequency correlations and controls}\label{sec:odd}

We evaluate the anomalous block $F(i\omega)=[(i\omega I-H_{\rm BdG})^{-1}]_{eh}$ and its odd-frequency part $F_o=[F(i\omega)-F(-i\omega)]/2$. The spatial diagnostic is the root-mean-square (RMS) of specified on-site or bond components, normalized by the number of pairs. Thus $M_0$ denotes the on-site RMS, not a complex spatial average. Appendix~\ref{app:odd} defines the real-space Nambu basis, Pauli relation, windows, complete spectral subspaces and numerical resolution rule. All results in this subsection are finite floating-point comparisons at $\omega/t\ge10^{-3}$, with $t=\Dz=\Dd=1$ and $\mu=\lambda=0$ before adding a stated texture or perturbation.

\paragraph{Geometry and holonomy.} At $\omega/t=10^{-3}$, the periodic $24\times24$ Klein bottle has $tM_0=7.8937853$ in the six-row seam band, versus $4.3091\times10^{-4}$ in the bulk band. The cut cylinder has $tM_0=8.7755510$ in the pooled edge band. Both signals are resolved; the working errors in the frozen reference calculation of these dimensionless RMS values are about $1.4\times10^{-6}$. Equal low-energy channel counts therefore do not imply equal anomalous amplitudes. Switching to the antiperiodic $x$ holonomy gives seam/edge values $2.7939\times10^{-4}$ and $3.0934\times10^{-4}$, respectively. For the Klein bottle this accompanies a change of the lowest $|E|/t$ from $2.9504\times10^{-6}$ to $0.2610524$. The periodic--antiperiodic on-site differences satisfy the registered comparison rule at the 17 common standard frequencies from $\omega/t=0.001$ to $10^{0.2}\simeq1.584893$.

An ordinary $24\times48$ torus with two diagonal-pairing sign walls, separated by $24$ rows, has $tM_0=2.9485876$ at each wall at the same frequency. Each wall is separately normalized. This nonzero signal precludes identifying Klein gluing by the mere presence of odd-frequency correlations. Nor is the Klein response larger in every channel: its interior vertical-bond RMS is $4.9954806/t$, below the ordinary-wall value $6.7737742/t$. Total size, local pairing texture and global connections differ between these geometries; their amplitude difference is not isolated evidence for a non-orientability effect. The uniform torus also has resolved horizontal and diagonal background at $\omega/t=0.1$, with $tM=0.00124637$ and $0.00172437$. Its on-site signal is unresolved by the present working error, rather than established to be exactly zero.

\begin{figure*}[!t]
\centering\includegraphics[width=0.96\textwidth]{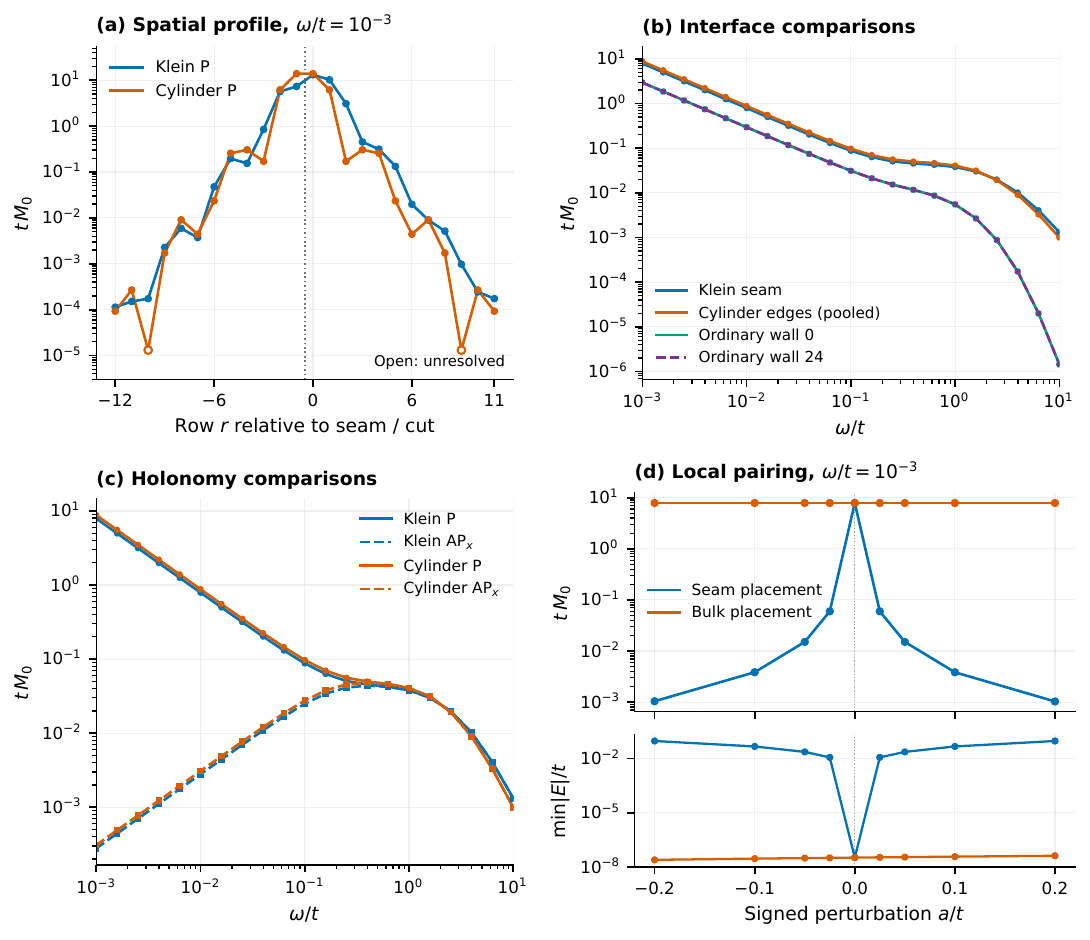}
\input{odd_figure_caption}
\label{fig:odd}
\end{figure*}

\paragraph{Local splitting and spectral attribution.} On $24\times32$, add $a(-1)^x$ to the vertical pairing on the source row $y=0$, or place the same term at $y=16$ as a bulk control. The registered amplitudes are $a/t=\pm0.025,\pm0.05,\pm0.1,\pm0.2$ (Fig.~\ref{fig:odd}). For $a=+0.1t$ at the seam, the lowest $|E|/t$ rises from $3.3849\times10^{-8}$ to $0.0458720$, while $tM_0$ at $\omega/t=10^{-3}$ falls from $7.8938540$ to $0.00383682$. The same bulk placement gives $|E|/t=3.8144\times10^{-8}$; its seam on-site RMS has no resolved difference from the clean case under the registered consecutive-point criterion. This statement concerns that diagnostic and window, not the absence of all bulk-perturbation effects.

For the periodic $24\times24$ Klein bottle at $\omega/t=10^{-3}$, the cutoff $|E|<0.1t$ selects two BdG eigenstates, the particle--hole partners. Their on-site seam RMS is $7.8936964/t$, while the remaining spectrum gives $9.00254\times10^{-5}/t$; the interference contribution to the squared RMS is $0.0014034144/t^2$. The complete lowest pair has mean projector weight $0.96772796$ in the seam band. These quantities support a low-energy attribution at this frequency. RMS contributions do not add linearly, and their ratio is not a probability. The finite-pair factor $-i\omega/(\epsilon^2+\omega^2)$ explains why both splitting and spectral weight matter (Appendix~\ref{app:odd}); no exact zero mode, fitted pole, or new protection mechanism is inferred. The supplement retains all four channels, both perturbation signs, two cutoffs and the lower-frequency resolution limits.

%% file: odd_figure_caption.tex
\caption{Finite odd-frequency correlations in the fixed pairing background, with $t=\Delta_0=\operatorname{Re}\Delta_{\mathrm d}=1$ and $\mu=\lambda=\operatorname{Im}\Delta_{\mathrm d}=0$.
$M_0$ is the root-mean-square onsite odd-frequency amplitude per site, in units of $t^{-1}$.
(a) Auxiliary row profiles at $\omega/t=10^{-3}$ for periodic Klein and cut-cylinder systems, both $24\times24$; the seam or cut lies between $r=-1$ and $0$. Open markers retain the two cylinder rows unresolved in the frozen reference calculation. Their original Accelerate flags differ; the shared-margin check in Appendix~\ref{app:odd-builds} keeps them unresolved.
(b) Periodic six-row interface averages for the Klein seam, pooled cylinder edges (three rows each), and each ordinary wall of the $24\times48$ two-wall torus W01, whose source-row pairing-sign texture differs from the earlier Bloch-wall control. The two ordinary-wall curves overlap at this scale; total size and local pairing templates differ from the Klein model.
(c) Periodic (P) and antiperiodic ($\mathrm{AP}_x$) holonomy in the same Klein and cylinder models.
Only standard frequencies $10^{-3}\leq\omega/t\leq10$ are displayed in (b,c).
(d) Seam onsite response (upper) and minimum absolute BdG energy (lower) for a $24\times32$ Klein system with signed staggered real vertical pairing on source row $0$ (seam) or $16$ (bulk); the clean point is shared.
All plotted values, errors and markers retain the frozen reference output. Error bars show working numerical estimates and are mostly smaller than the markers; for the spectrum they show the backward-error scale, not a rigorous interval. Lines guide the eye, without fits. Apparent overlap does not establish exact equality, and these finite correlations do not identify protected zero modes.}

%% file: odd_methods.tex
% WP5 draft; analytic identities and registered finite numerical methods.
% Input after \appendix. No additional packages or theorem counters required.
\section{Finite odd-frequency diagnostics}\label{app:odd}

\subsection{Nambu convention and exchange symmetry}\label{app:odd-def}
For $N_s=L_xL_y$ physical sites we use
$\Psi=(c_1,\ldots,c_{N_s},c_1^\dagger,\ldots,c_{N_s}^\dagger)^T$ and
\begin{equation}
 H_B=\begin{pmatrix}h&\Delta\\\Delta^\dagger&-h^T\end{pmatrix},
 \qquad h=h^\dagger,\quad\Delta=-\Delta^T.
 \label{eq:odd-nambu}
\end{equation}
Here $h=H_0-\mu I$. Sites are indexed by $i=xL_y+y$;
$x$ even/odd labels $A/B$. The real-space seam identifies
$(x,L_y)$ with $((1-x)\bmod L_x,0)$ and exchanges these labels.
The transpose in the hole block must be retained for complex hopping.
At real $\omega\ne0$, define
\begin{align}
 G(z)&=(zI-H_B)^{-1},\qquad F(z)=G_{eh}(z),\nonumber\\
 F_{o,e}(i\omega)&=\tfrac12[F(i\omega)\mp F(-i\omega)].
 \label{eq:odd-def}
\end{align}
All these Green functions have units $t^{-1}$.
The continuous imaginary-frequency grid below is not a finite-temperature
Matsubara data set: that interpretation would require a specified temperature
and $\omega_n=(2n+1)\pi T$.

The exchange relation follows directly from
$\tau_xH_B^T\tau_x=-H_B$. Indeed,
\begin{equation}
 \tau_xG^T(-z)\tau_x=(-zI+H_B)^{-1}=-G(z),
\end{equation}
so $F(z)=-F^T(-z)$, $F_o^T=F_o$, and $F_e^T=-F_e$.
Thus a nonzero spinless onsite anomalous correlation is necessarily odd in
frequency. This is an exchange constraint, not a topological test.
The equal-time Hamiltonian identity $c_i^\dagger c_i^\dagger=0$ does not forbid
a different-time onsite correlation. Nor does an odd-frequency correlation
in this fixed pairing background establish a self-consistent odd-frequency
order parameter or a measured conductance.

\subsection{Finite spectrum and the order of limits}\label{app:odd-spectrum}
For any finite Hermitian $H_B$, set
$G_{o,e}=[G(i\omega)\mp G(-i\omega)]/2$.
Multiplication of the two resolvents gives
\begin{align}
 G_o&=-i\omega(H_B^2+\omega^2I)^{-1},\nonumber\\
 G_e&=-H_B(H_B^2+\omega^2I)^{-1}.
 \label{eq:odd-resolvent}
\end{align}
Consequently, with $P_n$ the spectral projectors,
\begin{equation}
 F_o(i\omega)=-i\omega\sum_n\frac{(P_n)_{eh}}{E_n^2+\omega^2}.
 \label{eq:odd-spectral}
\end{equation}
This is an analytic identity. The implementation uses spectral sums or
linear solves, rather than explicitly squaring and inverting $H_B$.

For a nonzero pair $\pm\epsilon$, let $A=(P_+)_{eh}$.
Particle--hole conjugation gives $P_-=\tau_xP_+^*\tau_x$ and
$(P_-)_{eh}=A^T$. Therefore
\begin{align}
 F^{(\epsilon)}(z)&=\frac{A}{z-\epsilon}+\frac{A^T}{z+\epsilon},\nonumber\\
 F_o^{(\epsilon)}(i\omega)
 &=-\frac{i\omega(A+A^T)}{\epsilon^2+\omega^2}.
 \label{eq:odd-pair}
\end{align}
The same derivation applies to complete degenerate projectors. A contribution
can resemble $1/(i\omega)$ when $\epsilon\ll|\omega|$, yet is
$O(\omega)$ at fixed nonzero $\epsilon$ as $\omega\to0$; its particular
matrix residue can also vanish. If $H_B$ is invertible with
$g=\min_n|E_n|>0$, Eq.~\eqref{eq:odd-resolvent} implies
\begin{align}
 \|F_o(i\omega)\|_2&\le\frac{|\omega|}{g^2+\omega^2},\nonumber\\
 F_o(i\omega)&=-i\omega[H_B^{-2}]_{eh}+O(\omega^3/g^4).
 \label{eq:odd-limit}
\end{align}
An exact zero-energy projector instead contributes $(P_0)_{eh}/(i\omega)$,
whose selected entries need not be nonzero. Neither identity certifies that
a numerical near-zero eigenvalue is exactly zero or protected.
For a size-dependent pair,
$i\omega F_o^{(\epsilon_L)}=\omega^2(A_L+A_L^T)/(\omega^2+\epsilon_L^2)$;
taking $\omega\to0$ first and taking $\epsilon_L\to0$ first can differ.
No thermodynamic-limit inference is made from the present finite scan. We do not invoke spectral bulk--boundary correspondence \cite{TamuraHoshinoTanaka2019}: its anticommuting chiral operator and boundary trace have not been established for this Klein seam.

The two registered spectral cuts are $|E|<0.1t$ and $|E|<1.1t$.
Both retain complete particle--hole partners and degenerate spaces.
A cut within $50\eta$ of any computed $|E|$ is marked ambiguous, and its
decomposition is omitted rather than moved. None was ambiguous in this scan.
We retain $F_o=F_{o,\mathrm{low}}+F_{o,\mathrm{rest}}$, the complex entries
of both terms, and their interference. Rotating an exactly degenerate space
does not change its projector. A rotation within a space of unequal energies
must also rotate the restricted Hamiltonian in the resolvent; retaining
unchanged diagonal denominators would not be a valid basis change.

\subsection{Spatial averages and geometry controls}\label{app:odd-regions}
For a fixed set $B_\alpha(R)$ of sites ($\alpha=0$, $i=j$) or unordered
pairs, the reported quantity is
\begin{equation}
 M_\alpha(R,\omega)=\left[
 \frac{\sum_{(i,j)\in B_\alpha(R)}|(F_o)_{ij}|^2}{|B_\alpha(R)|}
 \right]^{1/2}.
 \label{eq:odd-rms}
\end{equation}
The other three channels have offsets $(1,0),(0,1),(1,1)$ in the local
lattice coordinates. Both endpoints must lie in $R$; $x$ wraps are retained,
whereas the common interior channels exclude $y$ wraps and crossings of
either ordinary wall. Interface-specific pairs are stored separately;
an absent cylinder interface is not assigned a zero correlation.
Under $H_B\mapsto W H_BW^\dagger$, $W=\operatorname{diag}(U,U^*)$ with
$U_{ij}=e^{i\chi_i}\delta_{ij}$, $F\mapsto UFU^T$.
Thus these fixed-pair RMS values are invariant under local basis phases,
although an unaligned average of complex amplitudes need not be.

The six-row seam band is $y=0,1,2,L_y-3,L_y-2,L_y-1$; the bulk band is
$y=L_y/2-3,\ldots,L_y/2+2$. Each three-row cylinder edge is also reported
separately, as well as their pooled six-row band. The uniform torus uses
the same labelled windows without a physical seam. For the $24\times48$
ordinary-wall torus, only $\Delta_{\rm d}$ changes sign, according to the
source row $y\ge24$. Its wall bands are $45,46,47,0,1,2$ and $21,\ldots,26$;
the corresponding bulk bands are $9,\ldots,14$ and $33,\ldots,38$.
Each wall is normalized separately. A six-row seam or wall band contains
144 onsite and 144 horizontal pairs, and 96 common interior vertical and
diagonal pairs; a six-row bulk band has 120 of each of the latter two.
All pair lists, source-sublattice groups and row profiles are retained.

Klein/cylinder comparisons change the cross-seam couplings, while
periodic/antiperiodic $x$ holonomy changes global transport without changing
the checked local plaquette flux. The ordinary-wall control matches the
24-row width of each chiral region and the averaging prescription; it
differs in total size, global connectivity and local pairing template.
Differences therefore cannot be assigned uniquely to non-orientability.

For any one of these component sets the normalized inner product gives
\begin{equation}
 \begin{split}
 M_{\rm full}^2={}&M_{\rm low}^2+M_{\rm rest}^2\\
       &+2\operatorname{Re}\langle F_{o,\rm low},F_{o,\rm rest}\rangle_R.
 \end{split}
 \label{eq:odd-interference}
\end{equation}
The RMS values are not additive and their ratio is not a probability.
Localization is instead measured by a complete low-energy projector,
using its particle-plus-hole probability in the region divided by its rank.

\subsection{Numerical checks and interpretation window}\label{app:odd-errors}
The scan fixes $t=\Delta_0=\operatorname{Re}\Delta_{\rm d}=1$ and
$\mu=\lambda=\operatorname{Im}\Delta_{\rm d}=0$. Its 23 configurations are
five $24\times24$ geometries/holonomies, a clean $24\times32$ Klein bottle,
eight seam and eight bulk perturbations of that bottle, and the ordinary-wall
torus above. The real vertical-pairing perturbation is
$\delta\Delta_{(x,y),(x,y+1)}=a(-1)^x$, with
$a/t=\pm0.025,\pm0.05,\pm0.1,\pm0.2$ and source row $y=0$ or $16$.
The standard grid is $\omega/t=10^{-8+0.2j}$, $j=0,\ldots,45$, with both
frequency signs evaluated. When $\hat g>100\eta$, gap-scaled frequencies
$\omega/\hat g=0.01,0.03,0.1,0.3,1,3,10,30,100$ are added only within
$10^{-10}\le\omega/t\le10$. There are 202 accepted additions and five
out-of-range omissions, giving 1,260 configuration--positive-frequency
records. These are deterministic evaluations, not independent samples.

Let $u$ denote the float64 machine epsilon. The registered working scale is
\begin{align}
 r_E&=\|H_BV-VE\|_F,\quad r_U=\|V^\dagger V-I\|_F,\nonumber\\
 \eta&=r_E+\|H_B\|_2r_U+10u\max(t,\|H_B\|_2).
 \label{eq:odd-eta}
\end{align}
The norm in this working estimate is evaluated as the largest absolute
computed Hermitian eigenvalue.
For each selected odd-frequency entry we use
\begin{align}
 B_{ij}^{\rm spec}&=\frac{\eta}{\omega^2}+\gamma_m S_{ij}+\frac{32u}{t},
 \quad\gamma_m=\frac{mu}{1-mu},\nonumber\\
 S_{ij}&=\sum_{n\,\mathrm{retained}}
 \frac{|\omega|\,|V_{e,in}|\,|V_{h,jn}|}{E_n^2+\omega^2}.
 \label{eq:odd-error}
\end{align}
Here $m=32(2N_s)+64$; the calculation stops if $mu\ge0.1$.
Each spectral subset uses its own sum but the full-matrix $\eta$.
The coefficients are preregistered working margins, not formally certified
roundoff bounds. In particular this odd-frequency estimate does not certify
the stored even-frequency controls.

Selected hole columns are also computed by direct linear solves at
$\omega/t=0.001,0.1,1$ for nine configurations, including the clean,
$a=+0.1t$ seam and bulk cases and both new controls. The columns include
$x=0,1$ and the first/third listed rows of each seam/bulk window, with both
walls and both bulk windows included for the wall control.
For $(i\omega I-H_B)X=B$, the exact-residual inequality
$\|\widehat X-X\|_2\le\|B-(i\omega I-H_B)\widehat X\|_2/|\omega|$
motivates the direct working estimate. We retain the larger Frobenius
residual from ordinary and compensated accumulation; the two signs give
$B^{\rm direct}=(B^++B^-)/2$ for the odd part.
The full-spectrum entry uses $\max(B^{\rm spec},B^{\rm direct})$ where
direct coverage exists and $B^{\rm spec}$ elsewhere. Low/rest entries use
their own spectral estimates. Floating-point residuals do not make these
interval certificates. The direct Pauli test covers the selected
$J\times J$ square, not the entire anomalous matrix.

Implementation controls include an analytic nonzero two-site bond
correlation, zero-pairing and equal-onsite-energy controls, complex hopping,
gauge covariance, degenerate-space rotation, structural residuals and
historical spectral anchors. Three fixed compensated-residual refinement
steps and a Rayleigh--Ritz refinement of the complete $|E|<0.1t$ space
were fixed during validation, before production. No correlation is
symmetrized and no small energy is set to zero. Comparison at validation
points uses componentwise $10^{-11}/t+5\times10^{-9}\max(|a|,|b|)$.
All 5,664 production checks passed, including seven applicable historical
spectral fields and 27 selected-column frequency checks.

The RMS working error is $b_R=[\operatorname{mean}_{R}B_{ij}^2]^{1/2}$.
A value is resolved when $M>10b_R$; a difference requires
$|M_A-M_B|>10(b_A+b_B)$ at three consecutive standard frequencies.
These rules are numerical resolution criteria, not statistical tests.
The manuscript's physical comparisons are restricted to
$0.001\le\omega/t\le10$. Lower-frequency data and resolution flags are
retained, but the additional direct checks needed for a low-frequency
interpretation were not performed. We make no power-law fit or numerical
zero-frequency-limit claim. The registered conditional width and trivial-phase
extensions were not activated: neither size scaling nor topological
specificity is claimed for these correlations.

\subsection{Reference build and classification sensitivity}\label{app:odd-builds}
The quoted working errors and resolution flags refer to the frozen
reference calculation. Its runtime library is OpenBLAS~0.3.29; the
OpenBLAS~0.3.21 string in NumPy's build metadata describes compilation,
not the library loaded at runtime. The package records the actual loaded
binary hashes, kernel and thread policy. A fresh installation with the
same Python and NumPy version labels but an Accelerate numerical backend
reproduces the physical values within the working comparison allowance,
while changing some error-dependent decisions. In the physical discussion
window, 21 of the 308 qualifying-run lists differ; eleven printed
remaining-spectrum U flags also differ. This failed cross-build
identical-classification test remains in the record.

A separately recorded, post hoc sensitivity policy uses the larger of the
two working errors at each matched standard-grid scalar,
$b_R^{\rm common}=\max(b_R^{\rm ref},b_R^{\rm alt})$.
We apply the original factor-ten and three-adjacent-point rules separately
to each run's unaveraged values, using this shared margin. In the tested
primary records the maximum is always the reference margin. The two
sets then agree for all 308 comparisons, all 47,840 standard-grid
full/low/rest primary scalar classifications and all 144 printed flags.
The corresponding lower-frequency agreement is only a diagnostic check;
it does not extend the physical interpretation window.

This is an empirical sensitivity check for the two recorded builds, not a
certified error bound or a guarantee for untested platforms. It does not
replace the original margins, promote the alternate-build-only signed
perturbation comparison to a physical claim, or erase the original failed
test. The supplement identifies the eleven sensitive table entries and
the reference provenance of figure markers. Near-zero printed values
already marked unresolved do not acquire meaningful last digits through
repetition across builds. Rebuilding the declared reference environment
and rerunning the original grid is a separate reproduction check.